\documentclass[preprint,a4paper]{elsarticle}

\usepackage[a4paper,margin=25mm]{geometry}
\usepackage[T1]{fontenc}
\usepackage[utf8]{inputenc}
\usepackage{mathptmx}
\usepackage{amsmath,amssymb,mathtools,bm}
\usepackage{graphicx}
\graphicspath{{figures/}}
\usepackage{booktabs,tabularx,array,longtable,adjustbox,multirow}
\usepackage{enumitem}
\usepackage{caption,subcaption,float,placeins}
\usepackage{xcolor}
\usepackage[hidelinks]{hyperref}
\usepackage{url}
\usepackage{algorithm}
\usepackage{algpseudocode}

\journal{}

\newcommand{\R}{\mathbb{R}}
\newcommand{\norm}[1]{\left\lVert#1\right\rVert}
\newcommand{\ind}{\mathbf{1}}
\newcommand{\method}[1]{\texttt{#1}}

\hypersetup{
  pdftitle={Toward Integrating Adaptive Experience Replay and Online Uncertainty Estimation in Safe Actor-Critic Optimal Control},
  pdfauthor={Mahshad Rastegarmoghaddam; Davoud Nikkhouy; Shima Samadzadeh},
  pdfsubject={}
}

\setlist[itemize]{leftmargin=*,itemsep=0.2em,topsep=0.3em}
\setlist[enumerate]{leftmargin=*,itemsep=0.25em,topsep=0.35em}
\begin{document}

\begin{frontmatter}

\title{Toward Integrating Adaptive Experience Replay and Online Uncertainty Estimation in Safe Actor-Critic Optimal Control}

\author[polimi]{Mahshad Rastegarmoghaddam\corref{cor1}}
\ead{mahshad.rastegarmoghaddam@mail.polimi.it}
\author[polimi]{Davoud Nikkhouy}
\ead{davoudnikkhoy@gmail.com}
\author[polimi]{Shima Samadzadeh}
\ead{shima.samadzade@gmail.com}
\cortext[cor1]{Corresponding author}
\affiliation[polimi]{organization={Department of Mechanical Engineering, Politecnico di Milano},
  city={Milan}, country={Italy}}

\begin{abstract}
Safe actor--critic control often treats barrier filtering, uncertainty estimation, and experience replay as separate modules, even though each changes the data used for learning and control. We develop an integrated architecture in which the uncertainty estimate updates the obstacle geometry used by a control barrier function, filter interventions and estimation residuals determine replay priority, and the critic learns from the executed rather than nominal action. We instantiate the architecture on a two-dimensional robot-navigation task with corrupted obstacle measurements and compare six component-matched configurations under common training budgets, random seeds, sensor streams, exploration, and disturbances. Evaluation includes a moderate post-training test, an eleven-level perception-noise sweep, and an exploratory extreme-stress test at multiplier $6.0$. In the extreme test, the integrated configuration recorded no contacts and reached the goal in all five evaluation seeds. Its mean cost was $7.63\pm0.44$ and its obstacle-belief root-mean-square error was $3.52\pm0.55$ cm. The uncertainty-estimation ablation also recorded no contacts but reached the goal in four of five seeds, with mean cost $8.96\pm2.08$ and belief error $11.08\pm1.23$ cm. A finite-training bound clarifies replay exposure, and a robust barrier condition states the required estimation-error and feasibility assumptions. The results support coupling estimation, safety filtering, and replay on this benchmark; broader safety and convergence claims require further study.
\end{abstract}

\begin{keyword}
Reinforcement learning \sep Actor-critic control \sep Control barrier functions \sep Experience replay \sep Online uncertainty estimation \sep Nonlinear optimal control \sep Safe optimal control
\end{keyword}

\end{frontmatter}

\section*{Data and code availability}
All numerical results were generated specifically for this study by executing the accompanying implementation on the common benchmark; none was copied from a prior publication. Code, deterministic seeds, data, figures, and protocol files are archived as version~\texttt{v1.1.0} at \url{https://doi.org/10.5281/zenodo.21515850}; the development repository is available at \url{https://github.com/SDNT8810/safe-actor-critic-aer-ue-reproducibility}.
For the reader’s convenience, the Supplementary Material has been appended to the end of this arXiv version. References to “Table S\#” and “Figure S\#” indicate the corresponding items in the Supplementary Material section.

\section{Introduction}

The control of nonlinear systems under state and input constraints remains a central problem in modern control, robotics, and intelligent decision-making. In many practical settings, the controller must optimize a performance criterion while simultaneously handling safety restrictions, partial model knowledge, exogenous disturbances, and limited data. Reinforcement learning (RL), especially in its actor-critic form, offers a flexible route toward approximate optimal control in such settings because it can combine value-function approximation, online policy improvement, and nonlinear function approximation within one computational pipeline \cite{sutton2018,li2023}. However, the successful deployment of RL-based controllers in safety-critical systems is still far from routine.

Two obstacles have become especially prominent. The first is safety during learning and execution. Standard actor-critic methods may generate transient actions that are locally useful from a reward viewpoint but unacceptable with respect to hard state or input constraints. The second is performance degradation under uncertainty. Even when a safe policy is found for a nominal model, the closed loop may deteriorate substantially if the environment contains unmodeled dynamics, time-varying disturbances, or distribution shifts. In addition, sample reuse---the mechanism that makes off-policy RL computationally attractive - is itself a double-edged sword: naive replay can introduce stale, redundant, or misleading information, especially near constraint boundaries.

Recent literature has addressed these issues from different angles. A strong line of work integrates control barrier functions (CBFs) into RL to create safety filters or safety-aware Bellman updates \cite{ames2017,marvi2021,jha2025,liu2023,hou2025,zhang2025}. A second line revisits replay design, emphasizing that the quality, diversity, and temporal relevance of retained transitions can strongly affect learning stability and efficiency \cite{adam2012,ERSurvey,mnih2015,schaul2015,gao2020}. A third line studies uncertainty-aware or disturbance-aware optimal control, often in continuous time, where recurrent networks, adaptive critics, and observer-based designs are used to estimate unknown dynamics online \cite{bian2022,wang2020,chen2022,wang2023,liu2025,elkenawy2020}. Yet these directions are still often developed in partial isolation.

This paper argues that safe actor-critic optimal control should be designed as an integrated architecture rather than as a sequence of loosely connected add-ons. In particular, replay and uncertainty estimation should not be treated as auxiliary implementation choices. They should be regarded as structural components that determine what the critic learns, what the safety layer observes, and how the policy responds near the safe-set boundary. This viewpoint is motivated by three recent lines of work: (i) safe actor-critic reinforcement learning with CBFs and experience replay for constrained optimal control problems \cite{hpv2025}; (ii) a bi-level architecture that combines a long short-term memory (LSTM) uncertainty estimator with an actor-critic policy for nonlinear systems with uncertain dynamics \cite{lstm2026}; and (iii) a self-organizing dual-buffer adaptive clustering experience replay mechanism for safe optimal control, with emphasis on memory efficiency and safety-critical replay selection \cite{sodacer2026}. The present work implements four interfaces: estimator-to-CBF geometry \cite{hpv2025}, estimator-to-replay priority \cite{lstm2026}, CBF-to-replay priority \cite{sodacer2026}, and executed-transition-to-critic learning. The central message is straightforward: safe actor-critic control, adaptive replay, and online uncertainty estimation should be co-designed. When this co-design principle is ignored, one often obtains good nominal learning but fragile safety, robust safety but poor sample efficiency, or accurate adaptation with insufficient replay diversity. When it is respected, a more balanced architecture becomes possible.

The main contributions of the present manuscript can be summarized as follows:
\begin{enumerate}
\item The paper synthesizes three research directions that are often treated separately in safe reinforcement learning for nonlinear optimal control: safety filtering, adaptive experience replay, and online uncertainty estimation.
\item It formulates an integrated architectural viewpoint in which the nominal actor-critic policy, the safety filter, the replay manager, and the uncertainty estimator are explicitly co-designed.
\item It clarifies the integration gap in the recent literature through a structured comparison of representative work strands.
\item It provides a compact design template for future methods, including an augmented replay representation, a modular learning loop, and a formal result on the relevance of boundary-critical samples.
\item It highlights the applicability of the proposed perspective across distinct domains, including robotics, autonomous systems, and epidemiological control.
\item It makes the template testable through explicit replay scores, a finite-training bound, a conditional robust-CBF inequality, and a six-configuration simulation study.
\end{enumerate}

The remainder of the paper is organized as follows. Section~\ref{sec:related} reviews the literature. Section~\ref{sec:problem} formulates the continuous-time constrained control problem. Section~\ref{sec:architecture} presents the integrated viewpoint and algorithmic template. Section~\ref{sec:experiments} reports the controlled numerical validation. Section~\ref{sec:discussion} discusses the evidence, limitations, and application bridges. Section~\ref{sec:conclusion} concludes the paper.

\section{Related Literature}
\label{sec:related}

The literature that motivates the present paper is scattered across several communities that do not always speak to one another directly. Control-theoretic studies emphasize stability, constraint satisfaction, and continuous-time structure; reinforcement-learning papers emphasize data efficiency and policy improvement; and application papers often focus on domain performance without isolating the architectural causes of success or failure. For that reason, the following review is organized around four recurring themes rather than around application areas alone. This organization makes it easier to see where the current literature already offers solid foundations and where a genuine integration problem still remains.

\subsection{Safe actor-critic optimal control}

A major advance in safe RL has been the incorporation of barrier-based conditions into policy learning or policy execution. CBF-based quadratic programs provide one rigorous way to ensure forward invariance of a safe set for safety-critical systems \cite{ames2017}. Building on this principle, safe RL methods have employed barrier inequalities either as online action-correction filters or as terms embedded directly in learning updates. Marvi and Kiumarsi \cite{marvi2021} formulated safe RL through CBF optimization, while more recent studies have extended the approach to off-policy learning, affine nonlinear systems with saturation, and transportation applications \cite{jha2025,liu2023,hou2025,zhang2025}. These works demonstrate that safety need not be treated as an external supervisory layer only; it can also be reflected in the value-update mechanism itself.

At the same time, the existing literature reveals a recurring trade-off. Strong safety filters protect the system but may distort the exploratory data on which the actor and critic are trained. If this distortion is ignored, the critic may learn an inconsistent value landscape. Conversely, if the actor is trained only on nominal transitions while the executed control is filtered, policy improvement may become unnecessarily slow. This observation already suggests that replay management must become safety-aware, since the buffer effectively determines what distribution the critic sees.

The application-oriented literature confirms the same point. In autonomous driving, safe soft actor-critic methods combined with disturbance-observer-based CBF filters improve robustness when the nominal vehicle model is inaccurate \cite{hou2025}. In powertrain control and motion planning, safe RL methods also benefit from explicit constrained optimal-control formulations \cite{hailemichael2023,zhang2025}. These contributions are valuable, but they also indicate that safety alone is not enough. Once the model is uncertain or the data are nonstationary, safety filtering must interact with adaptation and replay.

\subsection{Experience replay beyond uniform sampling}

Experience replay is one of the most consequential ideas in modern RL because it decouples sample collection from parameter updates and therefore improves both data efficiency and numerical stability \cite{adam2012,mnih2015}. Prioritized replay further refines this idea by sampling transitions with larger temporal-difference error more frequently \cite{schaul2015}. In practice, however, neither uniform replay nor simple priority scores are sufficient for safety-critical optimal control.

The reason is structural. In control tasks with state constraints, not all experiences carry equal information. Transitions near the safe-set boundary often contain disproportionately valuable information about the action-correction mechanism, barrier tightness, and feasibility margin. Likewise, transitions associated with abrupt disturbance changes can be more informative for uncertainty-aware policy adaptation than transitions collected in well-behaved nominal regimes. From this perspective, the replay buffer is not a passive data store; it is a compressed representation of the system's operational history.

Some application works already point in this direction. Hindsight experience replay has been shown to improve shared-control and robotic-manipulation settings by retaining task-relevant alternative outcomes \cite{gao2020}. Safe manipulation formulations under constrained Markov decision processes also show that task success and safe data efficiency must be handled jointly \cite{adjei2024}. Recent investigations of hybrid safe RL under distribution shift similarly underline the need for replay and uncertainty modeling to work together rather than separately \cite{hickman2025}. These studies motivate a more adaptive replay design, especially when the control task is nonlinear, continuous-time, and safety-constrained. Khalili-Amirabadi et al. \cite{sodacer2026} use fast and clustered slow buffers to preserve recency, diversity, and safety relevance. The present experiment instead uses a simpler finite mixed-priority buffer and does not claim to reproduce their complete algorithm.

\subsection{Online uncertainty estimation in continuous-time optimal control}

The literature on continuous-time adaptive critic control and approximate dynamic programming has long recognized that optimality and robustness cannot be separated cleanly \cite{bian2022,wang2020}. If disturbances or model mismatches are ignored, the critic approximates the wrong value function; if uncertainty estimation is inaccurate, policy updates can become either overly conservative or unstable.

Recent work has expanded this viewpoint. Echo-state and recurrent neural networks have been used for fault-tolerant and adaptive control, indicating that temporal memory can improve disturbance representation beyond static feedforward approximators \cite{chen2022,elkenawy2020}. Event-based and value-iteration-based robust adaptive critics continue this line by combining identification mechanisms with performance guarantees for nonaffine or disturbed systems \cite{wang2023,liu2025}. In parallel, application studies such as underwater-vehicle tracking demonstrate that recurrent architectures and attention mechanisms can be useful when disturbances evolve over time rather than appearing as white-noise perturbations \cite{tian2025}.

A relevant recent study is the bi-level LSTM-empowered actor-critic framework \cite{lstm2026}. There, the policy-optimization problem and the uncertainty-estimation problem are treated as two coupled layers. This separation improves interpretability and supports stability analysis under uncertain continuous-time dynamics. Conceptually, the importance of that result extends beyond the particular robot example. It indicates that uncertainty estimation should not be appended after policy training. It should shape the learning loop itself.

\subsection{Application bridges: robotics, autonomous systems, and health}

A useful test of any integrated perspective is whether it transfers across domains. Safe RL for navigation and cooperative manipulation shows that barrier-based safety and adaptive decision-making can be productive in robotics \cite{song2022,ganie2025}. At the same time, epidemiological control offers a quite different testbed, with slow state evolution, constrained interventions, and strong policy-safety interpretation. The HPV study by Khalili-Amirabadi et al. \cite{hpv2025} is important in this regard because it demonstrates that safe actor-critic ideas are not confined to mechanical systems; they can also support constrained public-health decision-making.

Taken together, the literature suggests a clear gap. Safety, replay design, and uncertainty estimation are all recognized as important, but a unifying control-oriented view is still missing. The next sections aim to articulate such a view.

\subsection{Recent event-triggered ADP and relevant CAEE studies}

Recent event-triggered adaptive dynamic programming (ADP) papers address adjacent but distinct questions. Hu et al. \cite{hu2026} combine integral RL, prescribed-performance transformation, and event triggering for multiplayer nonlinear games. Qin et al. \cite{qinmzs2026} use barrier transformation, a dynamic trigger, and critic-only experience replay for safe mixed zero-sum control. Related studies treat discontinuous state constraints \cite{qindiscontinuous2026} and actuator failures with full-state time-varying constraints \cite{qinfault2026}. These papers provide stronger plant-specific Lyapunov and event-triggering analyses than the present benchmark. The present contribution is complementary: it focuses on uncertainty-informed barrier geometry, executed-action critic learning, and realized finite-buffer safety/uncertainty replay. Direct numerical ranking would be invalid because the plants, objectives, communication assumptions, and constraint transformations differ.

Table~\ref{tab:related} summarizes the main strands of related work from the viewpoint adopted in this paper and highlights the still-open integration gap among safety filtering, replay design, and online uncertainty estimation.

\begin{table}[htbp]
\centering
\caption{Comparison of representative related-work strands. Existing studies typically emphasize safety, replay, or uncertainty adaptation in isolation; the present architecture co-designs the three components.}
\label{tab:related}
\scriptsize
\renewcommand{\arraystretch}{1.08}
\begin{adjustbox}{max width=\textwidth}
\begin{tabular}{p{2.8cm}p{2.3cm}p{2.5cm}p{2.6cm}p{1.7cm}p{5.1cm}}
\toprule
Work strand & Safety mechanism & Replay mechanism & Uncertainty estimator & Time setting & Main limitation \\
\midrule
Safe RL with CBFs \cite{marvi2021,jha2025,ames2017,liu2023} & CBF/QP safety filter & Usually standard replay or not central & Usually none or implicit nominal model & Both & Safety is explicit, but replay and uncertainty handling are only weakly integrated.\\
Replay-centric RL \cite{adam2012,mnih2015,schaul2015,gao2020} & Usually none & Uniform, PER, or HER & None & Mostly discrete & Data efficiency improves, but safety-critical and uncertainty-informative samples are not explicitly preserved.\\
Adaptive critic/robust ADP \cite{wang2020,wang2023,liu2025,bian2022} & Implicit robustness, rarely explicit safety filtering & Usually absent & Identification or robustness terms, but not modular replay-aware estimation & Continuous & Strong control-theoretic basis, but limited replay design and weak explicit safety integration.\\
Recurrent uncertainty-aware control \cite{elkenawy2020,chen2022,tian2025} & Application-dependent & Usually absent & RNN, echo-state network, attention, or LSTM & Continuous/\allowbreak mixed & Good temporal uncertainty modeling, but not tightly coupled with replay and safe actor-critic learning.\\
Application-level safe RL \cite{hpv2025,hou2025,adjei2024,song2022} & Domain-specific safety layer or explicit constraints & Standard or moderate replay use & Sometimes observer-based or application-specific & Mixed & Demonstrates practical value, but leaves the broader integration problem unresolved.\\
Event-triggered constrained ADP \cite{hu2026,qinmzs2026,qindiscontinuous2026,qinfault2026} & Barrier transformation or constrained control & Experience replay in selected designs & Plant/fault observers in selected designs & Continuous & Strong plant-specific stability and resource analysis; uncertainty-aware finite-buffer replay is not the central interface.\\
Related HPV component study \cite{hpv2025} & CBF-based safe actor-critic control & Experience replay & No dedicated recurrent uncertainty module & Continuous & Strong safety-oriented application bridge, but uncertainty estimation is not a first-class coupled module.\\
Related LSTM bi-level component study \cite{lstm2026} & Safe/robust control structure & Replay for actor-critic learning & LSTM-based online uncertainty estimator & Continuous & Strong uncertainty estimation, but replay is not explicitly safety- and diversity-aware.\\
Related SODACER component study \cite{sodacer2026} & Safe RL with CBFs & Dual-buffer clustered replay & No dedicated recurrent estimator & Continuous & Strong replay design, but uncertainty estimation is not integrated as a coupled module.\\
Present work & CBF-based safe action projection & Finite uniform, PER, or mixed adaptive replay & Online causal obstacle belief and scale & Continuous plant/discrete implementation & Evaluated on the stated two-dimensional benchmark; no general convergence or hardware claim.\\
\bottomrule
\end{tabular}
\end{adjustbox}
\end{table}

\section{Problem Formulation}
\label{sec:problem}

We consider a class of nonlinear continuous-time control systems
\begin{equation}
\dot{x}(t)=f(x(t))+g(x(t))u(t)+d(t),
\label{eq:dynamics}
\end{equation}
where $x(t)\in\R^n$ denotes the state, $u(t)\in\R^m$ is the control input, $f$ and $g$ are smooth dynamics, and $d(t)$ denotes a lumped uncertainty term that may include exogenous disturbances, parametric mismatch, unmodeled nonlinearities, and measurement-related effects. The control must satisfy admissibility bounds
\begin{equation}
u(t)\in\mathcal U:=\{u\in\R^m:|u_i|\le\bar u_i,\ i=1,\ldots,m\},
\label{eq:inputset}
\end{equation}
and the state must remain in a safe set
\begin{equation}
\mathcal X_{\mathrm{safe}}:=\{x\in\R^n:h_j(x)\ge0,\ j=1,\ldots,r\}.
\label{eq:safeset}
\end{equation}

The nominal optimal-control objective is to minimize an infinite-horizon performance index of the form
\begin{equation}
J(x_0,u)=\int_0^\infty e^{-\nu t}\ell(x(t),u(t))\,dt,
\label{eq:objective}
\end{equation}
where $\nu>0$ is a discount factor and $\ell(x,u)$ is a positive-definite running cost, often chosen as a weighted sum of tracking error and control effort. In continuous-time actor-critic formulations, the value function $V(x)$ satisfies a Hamilton-Jacobi-Bellman relation at least approximately, and the actor is updated using the critic gradient or a policy-gradient surrogate \cite{bian2022,wang2020}.

If safety were absent and $d(t)$ were known, one could attempt a standard approximate dynamic-programming or actor-critic construction. In the present setting, however, three complications must be addressed simultaneously:
\begin{enumerate}
\item the system should remain inside $\mathcal X_{\mathrm{safe}}$ during learning and execution;
\item the critic should learn from replayed data without being dominated by redundant or stale transitions; and
\item the uncertainty $d(t)$ should be estimated online well enough to preserve both performance and safety.
\end{enumerate}
These three requirements motivate the integrated architecture developed in the next section. For convenience, Table~\ref{tab:notation} summarizes the principal symbols that appear in the integrated architecture and the subsequent discussion.

\begin{table}[htbp]
\centering
\caption{Core notation used in the integrated safe actor-critic architecture.}
\label{tab:notation}
\small
\begin{tabular}{p{3.2cm}p{9.5cm}}
\toprule
Symbol & Meaning \\
\midrule
$x_t,x_{t+1}$ & Current and next states \\
$u_{a,t}$ & Nominal actor action before safety correction \\
$u_t$ & Executed safe action after safety filtering \\
$r_t$ & Instantaneous reward or cost feedback \\
$\hat d_t$ & Estimated lumped uncertainty \\
$\zeta_t$ & Recent trajectory window for uncertainty estimation \\
$\Pi_{\mathrm{safe}}$ & Safety filter based on barrier constraints \\
$V_\omega(x)$ & Critic value approximation \\
$\pi_\theta(x)$ & Actor policy \\
$\xi_t$ & Augmented replay transition \\
$s_t,e_t,n_t$ & Safety, uncertainty, and novelty scores for replay prioritization \\
$\delta_t$ & Temporal-difference error \\
$p_t$ & Replay priority \\
$\lambda_\bullet$ & Weights in mixed replay prioritization, where $\bullet$ represents TD, safe, unc, or nov \\
\bottomrule
\end{tabular}
\end{table}

\section{Toward an Integrated Architecture}
\label{sec:architecture}

The preceding discussion suggests that the next step is not merely to add another component to an existing safe RL pipeline, but to rethink the way the main modules exchange information. In particular, the replay buffer should not be treated as a passive storage unit, the uncertainty estimator should not be viewed as an optional add-on, and the safety filter should not be regarded as a purely post hoc correction layer. Instead, these modules should be designed as coupled elements of one learning architecture. The purpose of this section is therefore to articulate a compact design view that can support future algorithm development in a principled and application-independent manner.

\subsection{Design principle}

The key idea is to separate the learning system into four interacting modules:
\begin{enumerate}
\item a nominal actor-critic module that proposes control actions for approximate optimality;
\item a safety filter based on CBF inequalities that corrects unsafe actions;
\item an online uncertainty estimator that provides a running estimate $\hat d(t)$ of the lumped uncertainty; and
\item an adaptive replay manager that decides which transitions should be retained and how they should be resampled.
\end{enumerate}

The novelty of the viewpoint advocated here lies not in the mere identification of these components, but in the way their interactions are explicitly structured. The safety filter should use the uncertainty estimate, because barrier feasibility and future-state prediction depend on disturbance knowledge. The replay manager should use both the safety layer and the uncertainty estimator, because transitions close to barrier activation or with large estimation residuals are especially informative. The critic should be updated on the basis of the executed safe control, not merely the nominal actor output, because that is the action that truly determines the closed-loop trajectory.

\begin{figure}[htbp]
\centering
\includegraphics[width=0.98\textwidth]{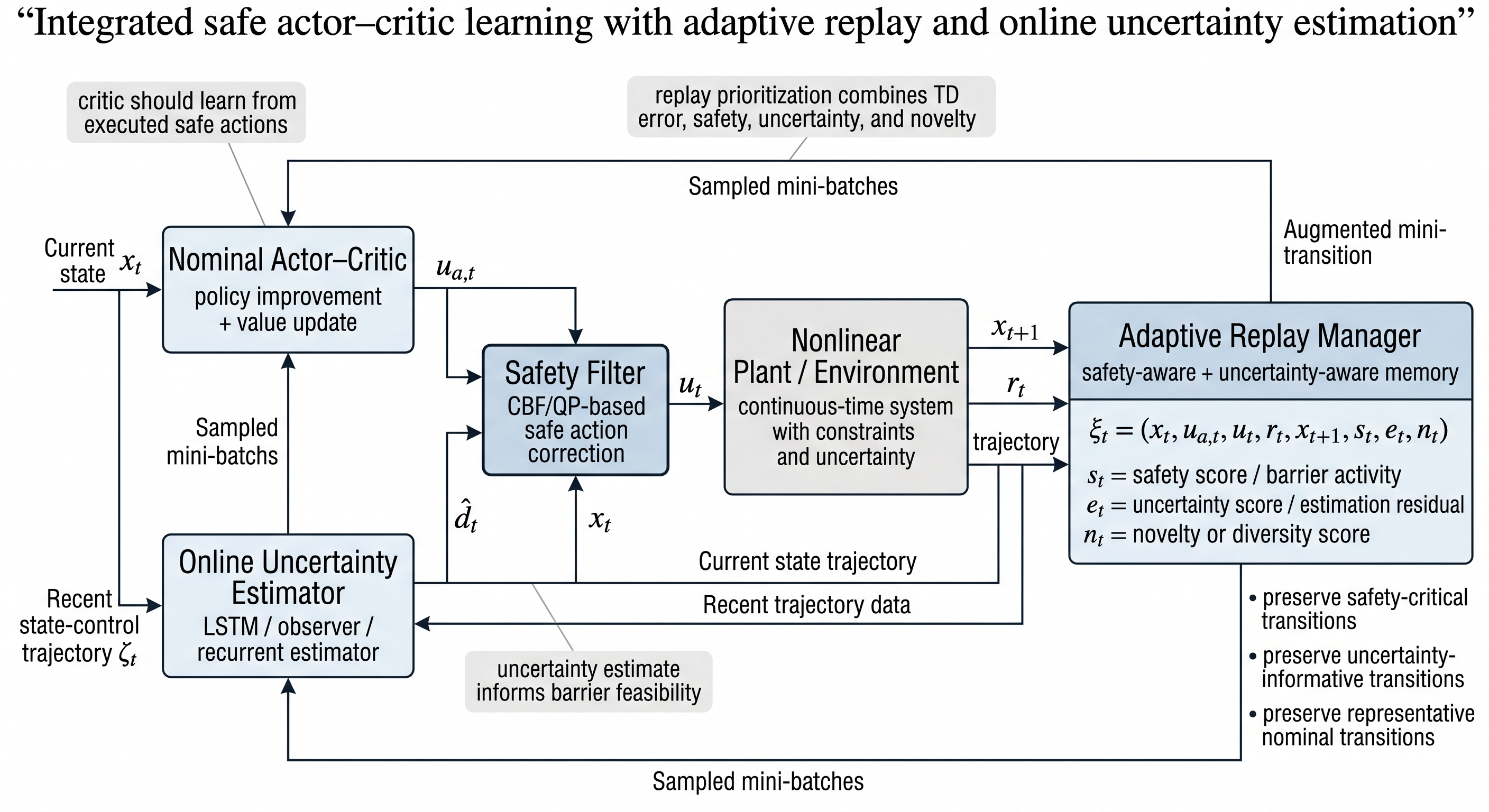}
\caption{Integrated four-module architecture for safe actor-critic optimal control with adaptive experience replay and online uncertainty estimation.}
\label{fig:architecture}
\end{figure}

Figure~\ref{fig:architecture} summarizes the integrated viewpoint advocated in this paper. Rather than treating safety filtering, replay design, and uncertainty estimation as loosely connected auxiliary mechanisms, the diagram emphasizes their mutual dependence within a single learning-and-control loop. In particular, the executed safe action, rather than the nominal actor output alone, determines the plant trajectory, the replay content, and ultimately the critic update. Likewise, the uncertainty estimator informs safety correction, while the replay manager preserves the transitions that are most informative for both safe control and adaptive learning.

In the generic formulation, $\hat d$ represents uncertainty in the plant dynamics. The experiment instantiates the same estimator-to-safety interface for perception uncertainty: a causal filter estimates static obstacle centers and their scales from detector data, these estimates define the barrier geometry, and plant disturbance is handled only through the fixed allowance $\bar d$. The experiment therefore does not implement online disturbance identification, an LSTM, or a clustered dual buffer. The controller uses only observable states, detector outputs, estimator beliefs, actions, filter activity, and contact events. True obstacle centers and clearances are reserved for simulation and evaluation; Supplementary Table~S2 gives the complete information contract and its invariance audit.

\subsection{Nominal actor-critic and safe action correction}

Let the critic approximate a value function $V_\omega(x)$ with parameters $\omega$, and let the nominal actor produce an action $u_a=\pi_\theta(x)$ with parameters $\theta$. The executed control is then obtained from a safety-correction stage,
\begin{equation}
u(t)=\Pi_{\mathrm{safe}}\bigl(x(t),u_a(t),\hat d(t)\bigr),
\label{eq:safefilter}
\end{equation}
where $\Pi_{\mathrm{safe}}$ denotes a filter that returns the closest admissible action satisfying the CBF constraints. In a standard formulation, the filter may solve a quadratic program of the form
\begin{equation}
\begin{aligned}
\min_{u\in\mathcal U}\quad & \norm{u-u_a}_R^2 \\
\text{s.t.}\quad & \nabla h_j(x)^\top\bigl(f(x)+g(x)u+\hat d(x)\bigr)
 +\alpha_j\bigl(h_j(x)\bigr)\ge0,
\end{aligned}
\label{eq:cbfqp}
\end{equation}
for all active safety constraints $j=1,\ldots,r$, where $R\succ0$ and $\alpha_j(\cdot)$ is an extended class-$\mathcal K$ function \cite{ames2017,marvi2021}. Equation~\eqref{eq:cbfqp} highlights a point that is sometimes understated in the RL literature: the safety filter is model-dependent. If the estimate $\hat d$ is poor, the filter may either become too conservative or fail to protect the true system adequately. Hence, uncertainty estimation and safety correction should be treated as coupled computations.

This coupling does not by itself guarantee recursive feasibility. Suppose $\norm{d(x)-\hat d(x)}\le\bar e_d$ and the error in the estimated barrier geometry contributes at most $\bar e_h$ to its derivative. A sufficient robust inequality is
\[
\nabla\hat h(x)^\top\bigl(f(x)+g(x)u+\hat d(x)\bigr)+\alpha(\hat h(x))
\ge \norm{\nabla\hat h(x)}\bar e_d+\bar e_h.
\]
If these bounds are valid and the robust admissible-action set is nonempty at every relevant state, the true derivative satisfies the intended CBF condition. If a bound is underestimated, the estimator is miscalibrated, the relative degree changes, or the input constraints make the robust set empty, forward invariance and recursive feasibility are not guaranteed. The benchmark filter below is therefore reported as an empirical robustification, not an unconditional certificate.

\subsection{Online uncertainty estimation}

The uncertainty estimator can be implemented in several ways, depending on the domain and computational budget. Observer-based adaptive critics, echo-state networks, diagonal recurrent neural networks, and LSTM architectures are all reasonable candidates \cite{chen2022,elkenawy2020,wang2023,tian2025}. In the present perspective, the estimator is updated online from recent state-control trajectories,
\begin{equation}
\hat d(t)=\mathcal E_\phi(\zeta_t),
\label{eq:estimator}
\end{equation}
where $\mathcal E_\phi$ is a parameterized estimator and $\zeta_t$ denotes a window of current and past states and controls. The LSTM-based bi-level framework in Khalili-Amirabadi et al. \cite{lstm2026} is especially relevant because it demonstrates how uncertainty estimation can be separated from, yet still coupled to, policy optimization.

For conceptual clarity, we recommend two design rules:
\begin{enumerate}
\item uncertainty estimation should be trained on transitions collected under the executed safe policy, not a purely nominal one; and
\item estimator confidence or residual magnitude should be made available to the replay manager.
\end{enumerate}
The second rule is important. A transition with a large temporal-difference error may be useful for critic training, but a transition with a small temporal-difference error and a very large uncertainty residual may still be crucial for model adaptation and future safety.

\subsection{Adaptive experience replay}

In standard replay, each transition is stored as $(x_t,u_t,r_t,x_{t+1})$. In safe actor-critic control, this tuple is often insufficient. We advocate storing an augmented tuple,
\begin{equation}
\xi_t=(x_t,u_{a,t},u_t,r_t,x_{t+1},\sigma_t,\rho_t,\eta_t),
\label{eq:tuple}
\end{equation}
where $u_{a,t}$ is the nominal actor action, $u_t$ is the executed safe action, $\sigma_t$ is a safety-related observable such as barrier slack or filter activation, $\rho_t$ is an uncertainty-related observable such as an estimation residual norm, and $\eta_t$ is a recency or context descriptor.

The experimental implementation also stores the exploration vector and active waypoint index required by its replayed actor update. Supplementary Table~S2 lists the complete controller-visible record, including filter activity, estimator innovation, detector validity, and observable jump scores; evaluator-only true geometry is excluded.

This augmented representation allows the replay manager to preserve three types of transitions:
\begin{enumerate}
\item safety-critical transitions, for which the CBF filter was active or nearly active;
\item uncertainty-informative transitions, for which the estimator residual was large or the operating regime changed abruptly; and
\item representative nominal transitions, which preserve coverage of the ordinary operating regime.
\end{enumerate}

This view naturally motivates dual-buffer or clustered replay. The SODACER work provides one implementation by combining a fast buffer for short-horizon adaptation with a clustered slow buffer for long-horizon diversity and redundancy control \cite{sodacer2026}. Conceptually, the clustered slow buffer is appealing because it compresses repeated nominal experiences and reserves memory for rare but structurally important transitions.

A generic sampling score may be written as
\begin{equation}
p_t=\lambda_{\mathrm{TD}}\bar\delta_t+\lambda_{\mathrm{safe}}s_t
+\lambda_{\mathrm{unc}}e_t+\lambda_{\mathrm{nov}}n_t,
\label{eq:genericpriority}
\end{equation}
where $\bar\delta_t$ is a normalized temporal-difference score, $s_t$ is a safety-importance score, $e_t$ is an uncertainty score, and $n_t$ is a novelty or diversity score. The weights $\lambda_\bullet$ are design parameters.

The implementation converts the raw observables $(\sigma_t,\rho_t,\eta_t)$ in Eq.~\eqref{eq:tuple} into bounded dimensionless scores in Eq.~\eqref{eq:genericpriority}:
\begin{align}
\bar\delta_t&=\min\left(1,\frac{|\delta_t|}{2.5}\right),\label{eq:tdscore}\\
s_t&=\min\Biggl(1,
e^{-\max(\tilde h_t,0)/0.075}
+0.34\ind_{\mathrm{CBF},t}\notag\\
&\hspace{3.1em}{}+0.30\min\left(1,\frac{\norm{u_t-u_{a,t}}}{0.11}\right)
+0.75\ind_{\mathrm{contact},t}\Biggr),\label{eq:safetyscore}\\
e_t&=\min\left(1,
\frac{0.58}{K}\sum_{i=1}^{K}\frac{\nu_{i,t}}{\max(2.5\sigma_{i,t},0.030)}\right.\notag\\
&\hspace{5.2em}\left.+0.30\bar j_t+0.28(1-\bar v_t)\right),\label{eq:uncscore}\\
n_t&=\min\left(1,
\frac{1}{\sqrt{1+N_t(\operatorname{cell}(x_t))}}
\right).\label{eq:novscore}
\end{align}
where $\tilde h_t$ is estimated barrier slack, $\nu_{i,t}$ is causal estimator innovation, $\bar j_t$ is the mean observable detector-jump score, $\bar v_t$ is the mean measurement-validity indicator, and $N_t$ is the prior occupancy of the state cell. All distances in the denominators are in meters. The mixed weights are
\begin{equation}
(\lambda_{\mathrm{TD}},\lambda_{\mathrm{safe}},\lambda_{\mathrm{unc}},\lambda_{\mathrm{nov}})
=(0.18,0.39,0.30,0.13),
\label{eq:weights}
\end{equation}
and $10^{-5}$ is added to every priority. For a retained buffer $\mathcal D$, the executable sampler tempers the priority and samples a mini-batch without replacement. At each sequential draw it uses
\begin{equation}
\Pr(i\mid\mathcal D)=\frac{p_i^{\beta}}{\sum_{k\in\mathcal D}p_k^{\beta}},
\qquad
\beta=0.78\ \text{for AER/Full},
\label{eq:sampleprob}
\end{equation}
with the denominator recomputed over the not-yet-selected entries. PER uses temporal-difference priority alone and $\beta=0.68$, whereas the uniform methods sample without replacement with equal probability. Thus Eq.~\eqref{eq:genericpriority} is not left as an unnormalized heuristic. The values are fixed engineering choices, not claimed global optima; the Supplementary Material reports a post-hoc one-at-a-time weight sensitivity diagnostic and does not use it to select the reported configuration.

\textbf{Proposition 1 (finite-training exposure of a retained critical transition).}
Consider a finite replay buffer and $M$ subsequent replay draws. Let $i$ be a safety-critical transition that remains stored during those draws. If its conditional probability of selection at draw $m$, given the complete history $\mathcal F_{m-1}$, satisfies
\begin{equation}
\Pr(i\text{ selected at draw }m\mid\mathcal F_{m-1})\ge p_{\min}>0,
\label{eq:pmin}
\end{equation}
then
\begin{equation}
\Pr(i\text{ is never selected in the next }M\text{ draws})
\le(1-p_{\min})^M\le e^{-Mp_{\min}},
\label{eq:neverdrawn}
\end{equation}
and consequently
\begin{equation}
\Pr(i\text{ is selected at least once})\ge1-e^{-Mp_{\min}}.
\label{eq:atleastonce}
\end{equation}
If the total conditional sampling mass assigned to the currently retained critical set $\mathcal C$ is at least $q_{\min}>0$ on every draw, then the probability that no critical transition is sampled in $M$ draws is at most $e^{-Mq_{\min}}$, and the expected number of critical draws is at least $Mq_{\min}$.

\textit{Proof.} By the chain rule,
\begin{align}
\Pr(i\text{ never selected})
&=\prod_{m=1}^{M}\Pr(i\text{ not selected at }m\mid i\text{ was not selected earlier})\\
&\le(1-p_{\min})^M\le e^{-Mp_{\min}}.
\end{align}
The critical-set statement follows by replacing the individual event with selection from $\mathcal C$. If $Z_m$ indicates a critical draw, then $\mathbb E[Z_m\mid\mathcal F_{m-1}]\ge q_{\min}$, and linearity of expectation gives $\mathbb E[\sum_{m=1}^{M}Z_m]\ge Mq_{\min}$. \hfill$\square$

The retained-transition condition is essential. Proposition~1 is a replay-exposure result, not an eviction guarantee, actor-critic convergence theorem, or closed-loop invariance theorem. This is why the experiment audits generated, retained, and sampled event fractions separately.

For the implemented $N=360$ buffer, the priority floor gives the deliberately loose bound $p_{\min}\ge3.51\times10^{-7}$ and only $1-e^{-Mp_{\min}}\approx0.00123$ over $M=3520$ draw positions. This $M$ is the maximum available across a full run for a transition inserted before the first replay update and retained throughout; later arrivals have fewer opportunities. We therefore assess practical coverage using the observed generation, retention, and sampling fractions in Supplementary Table~S10 rather than this worst-case floor.

\subsection{Integrated learning loop}

Algorithm~\ref{alg:integrated} summarizes the proposed perspective.

\begin{algorithm}[H]
\caption{Integrated safe actor-critic learning with adaptive replay and uncertainty estimation}
\label{alg:integrated}
\begin{algorithmic}[1]
\State Initialize critic parameters $\omega$, actor parameters $\theta$, estimator parameters $\phi$, and replay memory $\mathcal D$.
\For{each interaction step $t$}
  \State Observe state $x_t$.
  \State Compute nominal action $u_{a,t}=\pi_\theta(x_t)$.
  \State Estimate uncertainty $\hat d_t=\mathcal E_\phi(\zeta_t)$.
  \State Compute safe executed action $u_t=\Pi_{\mathrm{safe}}(x_t,u_{a,t},\hat d_t)$.
  \State Apply $u_t$; observe reward $r_t$ and next state $x_{t+1}$.
  \State Compute safety score $s_t$, uncertainty score $e_t$, and novelty score $n_t$.
  \State Store augmented transition $\xi_t$ in replay memory $\mathcal D$.
  \State Sample a mini-batch using a mixed priority based on TD, safety, uncertainty, and diversity.
  \State Update the critic using executed safe transitions.
  \State Update the actor using the current critic and safety-aware targets.
  \State Update the estimator using recent trajectory windows or replayed uncertainty-informative samples.
  \State Periodically compress, cluster, or rebalance the replay memory when the chosen replay design requires it.
\EndFor
\end{algorithmic}
\end{algorithm}

In the experiment, belief and action updates occur at every step, whereas actor--critic replay updates occur after each episode. The critic is fitted to realized costs and next states following the executed safe action. The waypoint actor uses the critic's clipped temporal-difference residual and stored exploration draw in a replay-weighted score-function update. Because no importance ratios are used, this is a biased off-policy surrogate; no isolated gain from the learned value baseline is claimed.

\subsection{Why integration matters}
\label{sec:whyintegration}

The expected advantages of the integrated architecture can be summarized in four points:
\begin{enumerate}
\item \textbf{Critic consistency.} If the critic is updated using executed safe actions and safety-aware replay, then the learned value function is more closely aligned with the true closed-loop behavior than in a nominal-only training setup. This reduces the mismatch between the policy that is proposed and the policy that is actually executed.
\item \textbf{Informative replay.} Adaptive replay can preserve rare but highly informative transitions associated with barrier activation, abrupt disturbance changes, or regime shifts. These are precisely the transitions that standard uniform replay tends to under-represent.
\item \textbf{Reliable safety filtering.} If the safety filter receives online uncertainty estimates rather than relying on a fixed nominal model, then barrier constraints can better reflect the true local dynamics. This improves the practical reliability of action correction under model mismatch.
\item \textbf{Cross-domain transferability.} Because the architecture is modular, it can inform designs across applications with very different physical interpretations. In robotics, $d(t)$ may represent slip, contact variation, or actuator uncertainty; in epidemiological control, it may reflect time-varying transmission, compliance variation, or intervention mismatch.
\end{enumerate}

Taken together, these four points explain why the proposed perspective is more than a simple aggregation of existing ingredients. Its value lies in making the interactions among learning, safety, replay, and uncertainty estimation explicit at the architectural level.

\begin{figure}[htbp]
\centering
\includegraphics[width=0.5\textwidth]{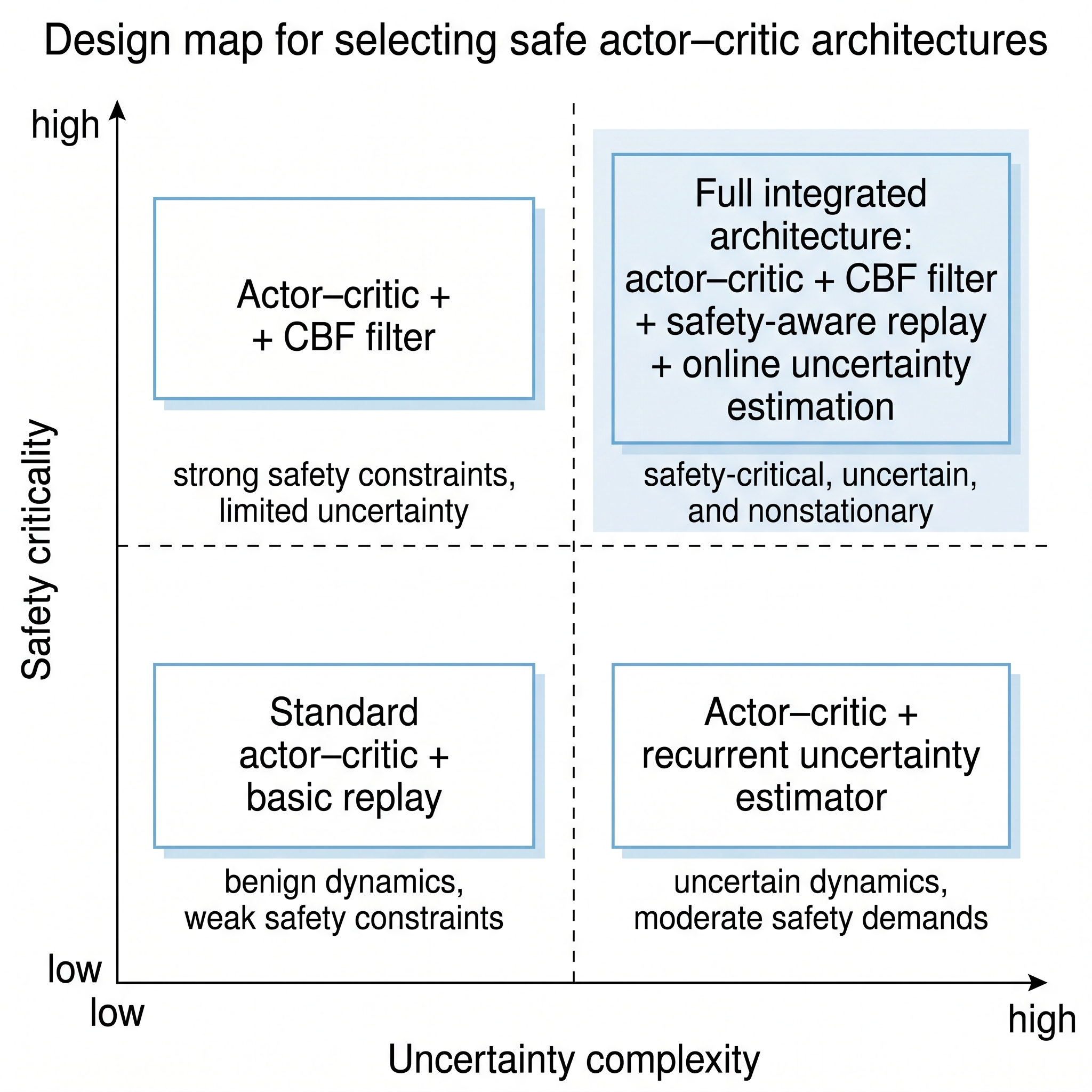}
\caption{A design map for selecting an appropriate safe actor-critic architecture based on uncertainty complexity and safety criticality. The upper-right quadrant corresponds to the main focus of this paper, namely settings in which safety-aware replay and online uncertainty estimation should be co-designed with the nominal actor-critic policy and the safety filter.}
\label{fig:designmap}
\end{figure}

Figure~\ref{fig:designmap} provides a compact design map that situates the proposed perspective within a broader class of safe actor-critic architectures. The figure is not intended as a rigid taxonomy, but as a practical guideline: as uncertainty becomes more structured and safety requirements become more stringent, the architecture must evolve from a basic actor-critic controller toward a fully integrated design in which safety filtering, replay management, and online uncertainty estimation are explicitly co-designed.

\begin{figure}[htbp]
\centering
\includegraphics[width=0.95\textwidth]{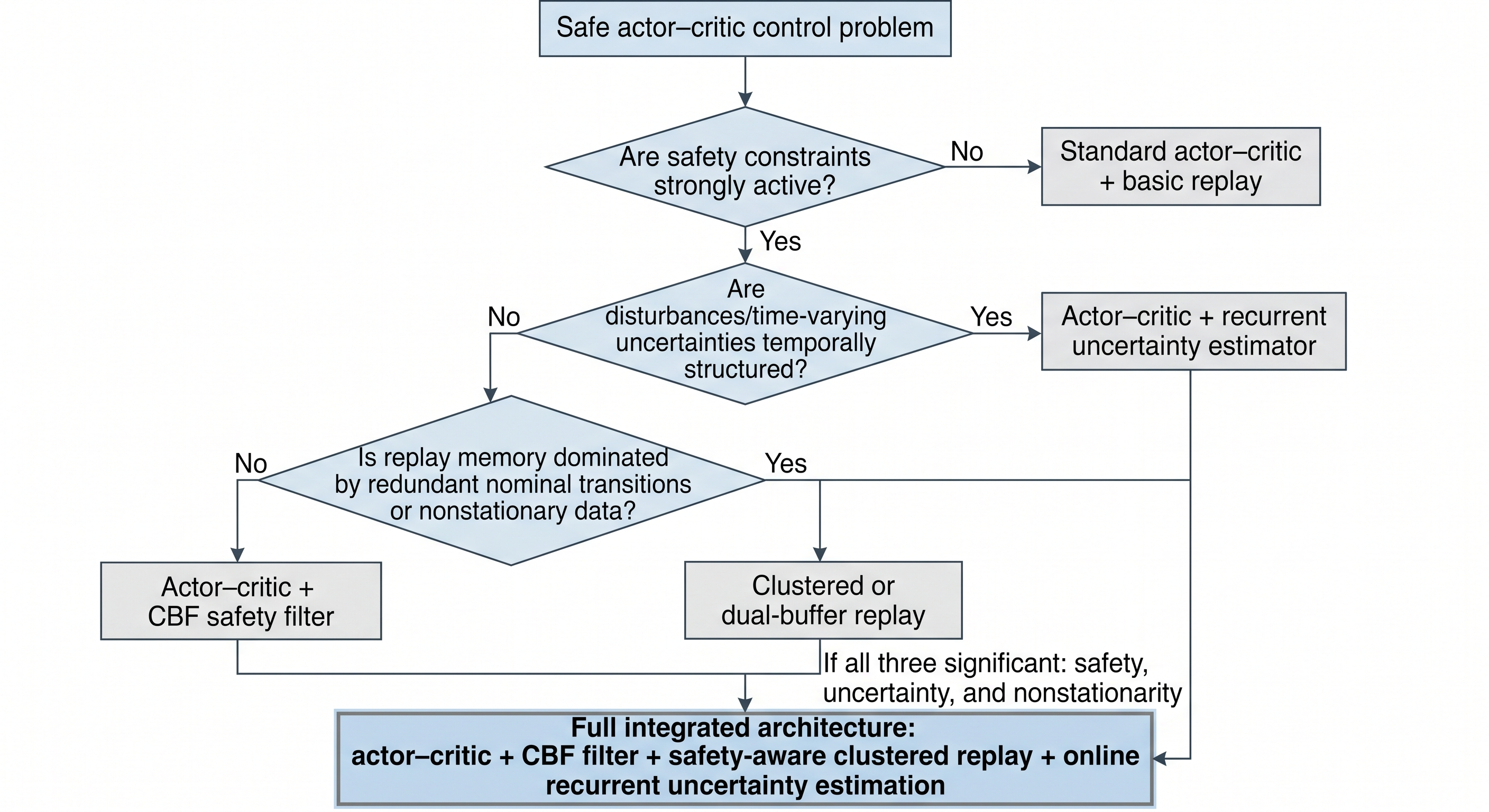}
\caption{Practical guide for selecting replay and uncertainty-estimation components.}
\label{fig:designguideline}
\end{figure}

Figure~\ref{fig:designguideline} summarizes when basic replay is sufficient and when safety, structured uncertainty, and replay nonstationarity motivate a more integrated design. The complementary design taxonomy is retained as Supplementary Figure~S8.

Figure~\ref{fig:designguideline} and Supplementary Figure~S8 are conceptual guides rather than descriptions of the implemented estimator and buffer. The static-center filter used below is appropriate only for physically static obstacles.

\FloatBarrier

\section{Experimental Validation}
\label{sec:experiments}

\subsection{Benchmark, observation process, and data partition}

The benchmark is a $3.20\,\mathrm{m}\times3.20\,\mathrm{m}$ cluttered workspace with known rectangular obstacles and two hidden-center circular obstacles. The robot radius is $0.062\,\mathrm{m}$, the continuous planar velocity is limited to $0.72\,\mathrm{m/s}$, and the integration step is $0.055\,\mathrm{s}$. The start and goal are $(0.22,2.82)\,\mathrm{m}$ and $(1.62,1.58)\,\mathrm{m}$. The two uncertain obstacle centers are $(0.82,1.60)\,\mathrm{m}$ and $(1.82,0.92)\,\mathrm{m}$, with radii $0.070\,\mathrm{m}$ and $0.100\,\mathrm{m}$. These physical obstacles remain static.

For uncertain obstacle $i$, the common detector reports
\begin{equation}
z_i(t)=c_i+b_i+w_i^{\mathrm{corr}}(t)+v_i(t)+d_i^{\mathrm{sin}}(t)+j_i(t),
\label{eq:sensor}
\end{equation}
when a measurement is valid. Here $c_i$ is the hidden static center, $b_i$ is episode-level registration bias, $w_i^{\mathrm{corr}}$ is colored drift, $v_i$ is white jitter, $d_i^{\mathrm{sin}}$ is smooth drift, and $j_i$ is a temporary localization jump. During an intermittent hold, the prior report is repeated and the validity bit is set to zero. For a fixed seed, episode identifier, and severity, every method receives the same arrays of detector centers, validity bits, observable jump scores, exploration draws, and plant disturbance.

Each method is trained for 10 episodes of 115 transitions and evaluated, without further learning, for 280 transitions over five paired seeds. Goal success is entry into a $5.5$ cm tolerance. The procedurally generated experiment has a training partition but no independent validation set. Evaluation comprises a moderate multiplier $2.2$, an eleven-level perception-noise sweep, and an exploratory multiplier-$6.0$ test selected after the initial analysis. Supplementary Tables~S1--S3 give the complete protocol and parameters.

At multiplier $6.0$, detector bias is $6.0$--$11.4$ cm, smooth drift reaches $5.4$ cm, temporary jumps begin at $30$--$42$ cm, and intermittent holds repeat stale measurements. The detector nevertheless reports a $1$ cm standard deviation, representing overconfidence under distribution shift. The plant-disturbance multiplier remains $1.0$; Supplementary Table~S1 gives the full timing and noise model.

The simulated plant and evaluation objective are fully discrete and use the executed post-filter velocity:
\begin{align}
p_{t+1}&=\operatorname{clip}_{\Omega}\!\left[p_t+0.055\,(u_t+d_t)\right],
\qquad \norm{u_t}\le0.72,\label{eq:plantupdate}\\
\ell_t&=0.036\norm{p_t-g}^{2}+0.020\norm{u_t}^{2}
+0.012\norm{u_t-u_{a,t}}^{2}+0.036\norm{u_t-u_{a,t}}\notag\\
&\quad{}+5.2\ind_{\mathrm{contact},t}
+\max\!\left(0,0.010-h_t^{\mathrm{known}}\right)^{2}.
\label{eq:benchmarkcost}
\end{align}
Here $d_t$ is the shared exogenous velocity disturbance, $\Omega$ clips the robot center to the workspace after accounting for its radius, $g$ is the goal, and $h_t^{\mathrm{known}}$ is signed clearance from the known map. The contact indicator is an observable bumper event; hidden true geometry is used to generate and evaluate contact but is not stored as a continuous learning signal. Supplementary Tables~S1--S3 give the complete protocol, stored-field schema, waypoint/corridor constraints, and all actor, critic, potential-field, estimator, CBF, and buffer constants.

\subsection{Progressive comparison configurations and common tuning budget}

The comparison is progressive and component-matched. \method{AC} is the common waypoint actor--critic with uniform replay. \method{AC+CBF} adds CBF projection \cite{ames2017,marvi2021}; \method{AC+CBF+PER} adds temporal-difference prioritized replay \cite{schaul2015}; \method{AC+CBF+UE} adds online uncertainty estimation \cite{hou2025,lstm2026}; and \method{AC+CBF+AER} adds adaptive safety-aware replay \cite{sodacer2026}. \method{Full} combines the estimator and adaptive replay with the safety filter. These are controlled in-house configurations, not copied results or exact reproductions of the cited methods, whose plants and objectives differ.

\begin{table}[htbp]
\centering
\caption{Progressive component-wise comparison. All configurations share the plant, cost, actor--critic update, training budget, seeds, and sensor, exploration, and disturbance streams. References identify the motivating literature strand.}
\label{tab:methods}
\small
\begin{tabularx}{\textwidth}{lccccX}
\toprule
Method & CBF & PER & UE & AER & Description \\
\midrule
\method{AC} \cite{sutton2018} & no & no & no & no & Waypoint actor--critic with a replay-weighted TD-residual score-function surrogate and uniform replay.\\
\method{AC+CBF} \cite{ames2017,marvi2021} & yes & no & no & no & CBF uses the common raw or held detector center.\\
\method{AC+CBF+PER} \cite{schaul2015} & yes & yes & no & no & Temporal-difference priority only.\\
\method{AC+CBF+UE} \cite{lstm2026} & yes & no & yes & no & Robust online static-center filter with uniform replay.\\
\method{AC+CBF+AER} \cite{sodacer2026} & yes & no & no & yes & Mixed adaptive replay; the CBF still uses the raw or held center.\\
\method{Full} (present) & yes & no & yes & yes & Online filter, replay-calibrated static anchor, robust CBF, and mixed replay.\\
\bottomrule
\end{tabularx}
\end{table}

All configurations use the same 10 training episodes, buffer capacity 360, batch size 44, eight replay updates per episode, and five seeds. No method-specific hyperparameter search or independent validation set was used; the common hand-fixed values are listed in Supplementary Table~S3. The post-hoc $\pm20\%$ replay-weight check in Table~S4 was not used to choose the reported configuration. This controls differential tuning but does not make every ablation individually optimal.

The estimator produces a causal center belief $\hat c_{i,t}$, scale $\sigma_{i,t}$, and innovation $\nu_{i,t}=\norm{z_{i,t}-\hat c_{i,t|t-1}}$; invalid measurements receive zero gain, and no true center label is used. For a circular obstacle, the filter uses
\begin{equation}
\hat h_i(p,t)=\norm{p-\hat c_i(t)}-\bigl(r_i+r_R+\rho_i(t)\bigr),
\qquad \rho_i(t)=0.009+1.20\sigma_i(t),
\label{eq:benchmarkbarrier}
\end{equation}
and projects the velocity onto active half-space approximations of
\begin{equation}
\nabla\hat h_i(p,t)^\top u\ge-\alpha\hat h_i(p,t)+\bar d+(0.18\,\mathrm{s}^{-1})\rho_i(t),
\label{eq:benchmarkcbf}
\end{equation}
subject to the input limits. Here $\bar d=0.012\,\mathrm{m/s}$ is a design allowance, not a hard stochastic bound, and $0.18\,\mathrm{s}^{-1}$ converts the distance envelope $\rho_i$ into a velocity margin. Known rectangles and workspace walls receive a discrete-time guard; the experiment therefore evaluates an empirical robustification rather than a certified stochastic guarantee.

\subsection{Metrics and statistical analysis}

Safety is evaluated against hidden true geometry. The signed clearance is
\begin{equation}
h_t=\min_i\left(\norm{p_t-c_i^{\mathrm{true}}}-r_i-r_R\right),
\label{eq:trueclearance}
\end{equation}
combined with distances to known rectangles and workspace walls. True clearance is evaluation-only. We report cost, violations, clearance, belief RMSE, goal performance, filter intervention, and runtime as mean $\pm$ sample standard deviation over five paired seeds. Friedman tests compare all configurations and paired Wilcoxon tests compare each ablation with Full; the $p$-values are unadjusted exploratory diagnostics. Supplementary Tables~S5--S10 and Figures~S1--S2 provide seed-level results, complete tests, replay fractions, and additional plots.

\subsection{Additional exploratory extreme-stress results}

Table~\ref{tab:severe} reports the multiplier-$6.0$ test. Safety, goal completion, cost, and estimation error are interpreted jointly.

\begin{table}[htbp]
\centering
\caption{Selected metrics from the extreme perception-stress test at multiplier $6.0$ (mean $\pm$ sample standard deviation over five seeds). Best means are bold; arrows indicate the preferred direction. Supplementary Table~S6 reports all metrics.}
\label{tab:severe}
\footnotesize
\begin{tabular}{lcccccc}
\toprule
Method & Cost $\downarrow$ & Viol. $\downarrow$ & Min. global $h$ $\uparrow$ & Uncertain clear. $\uparrow$ & Belief RMSE $\downarrow$ & Goal success $\uparrow$ \\
 & & (\%) & (cm) & (cm) & (cm) & (\%)\\
\midrule
\method{AC} \cite{sutton2018,li2023} & $46.99\pm19.28$ & $2.79\pm1.32$ & $-1.76\pm0.80$ & $-1.76\pm0.80$ & $16.71\pm1.55$ & $\mathbf{100.0\pm0.0}$\\
\method{AC+CBF} \cite{ames2017,marvi2021} & $31.00\pm12.43$ & $1.64\pm0.86$ & $-1.26\pm0.40$ & $-1.26\pm0.40$ & $16.71\pm1.55$ & $\mathbf{100.0\pm0.0}$\\
\method{AC+CBF+PER} \cite{schaul2015} & $39.08\pm24.08$ & $2.21\pm1.64$ & $-1.04\pm1.91$ & $-1.04\pm1.91$ & $16.71\pm1.55$ & $\mathbf{100.0\pm0.0}$\\
\method{AC+CBF+UE} \cite{lstm2026} & $8.96\pm2.08$ & $\mathbf{0.00\pm0.00}$ & $\mathbf{3.22\pm2.81}$ & $3.94\pm3.37$ & $11.08\pm1.23$ & $80.0\pm44.7$\\
\method{AC+CBF+AER} \cite{sodacer2026} & $22.62\pm12.82$ & $1.07\pm0.87$ & $-0.35\pm1.55$ & $-0.35\pm1.55$ & $16.71\pm1.55$ & $\mathbf{100.0\pm0.0}$\\
\method{Full} (present) & $\mathbf{7.63\pm0.44}$ & $\mathbf{0.00\pm0.00}$ & $2.71\pm0.23$ & $\mathbf{6.95\pm0.26}$ & $\mathbf{3.52\pm0.55}$ & $\mathbf{100.0\pm0.0}$\\
\bottomrule
\end{tabular}
\end{table}

UE and Full are the only configurations with zero contacts; Full reaches the goal in five of five seeds versus four of five for UE and has lower mean cost and belief error. Friedman tests detect differences in cost, violations, clearance, and belief RMSE ($p\le0.0045$), but the paired Full--UE tests do not reach $p<0.05$. The result is therefore a practically relevant joint trend on this benchmark, not proof of universal superiority. Complete seed-level and statistical results are in Supplementary Tables~S6--S9.

\begin{figure}[htbp]
\centering
\begin{subfigure}[t]{0.72\textwidth}
\includegraphics[width=\textwidth]{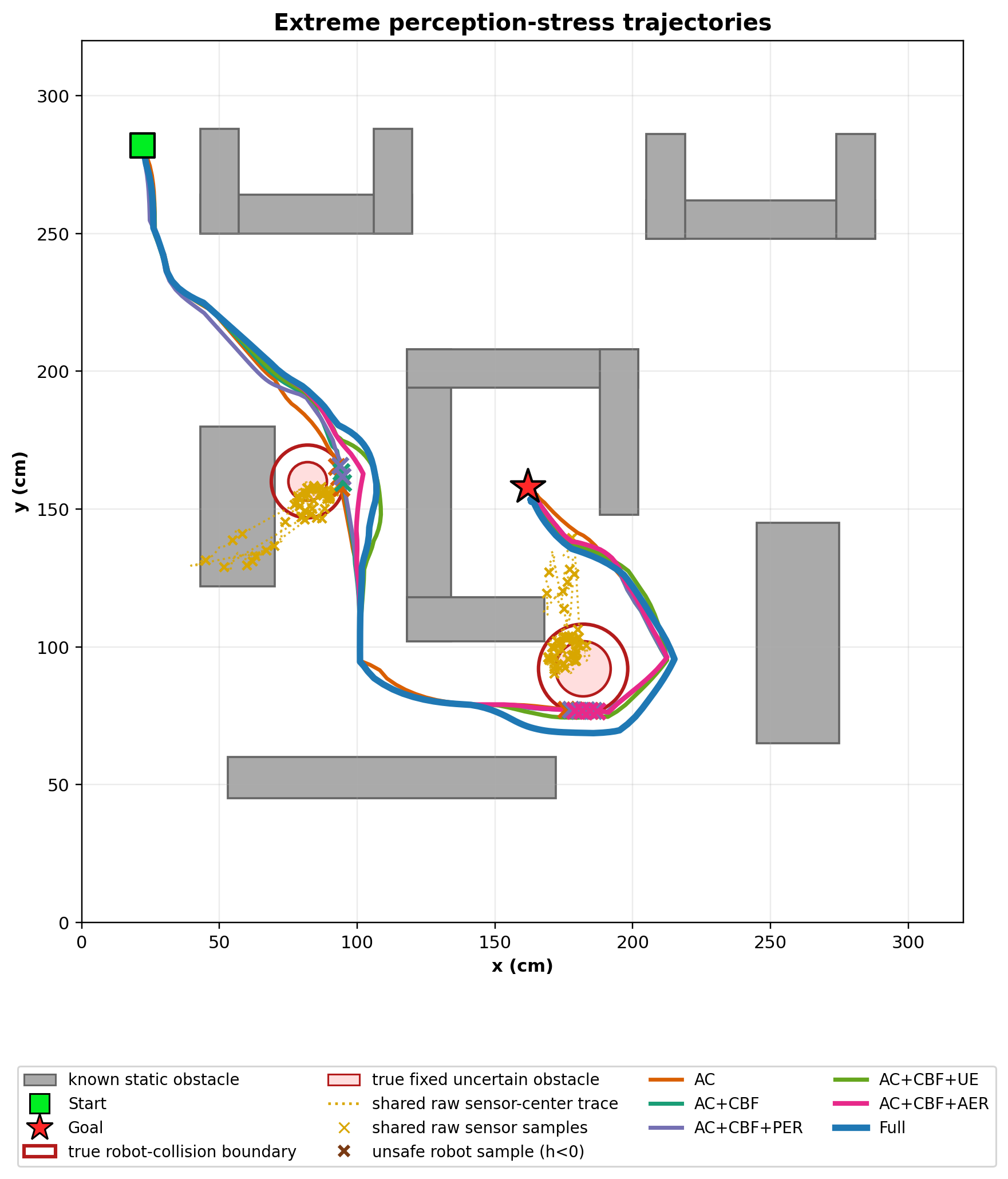}
\caption{Trajectories for visualization seed 4.}
\label{fig:traj}
\end{subfigure}\\[3pt]
\begin{subfigure}[t]{0.88\textwidth}
\includegraphics[width=\textwidth]{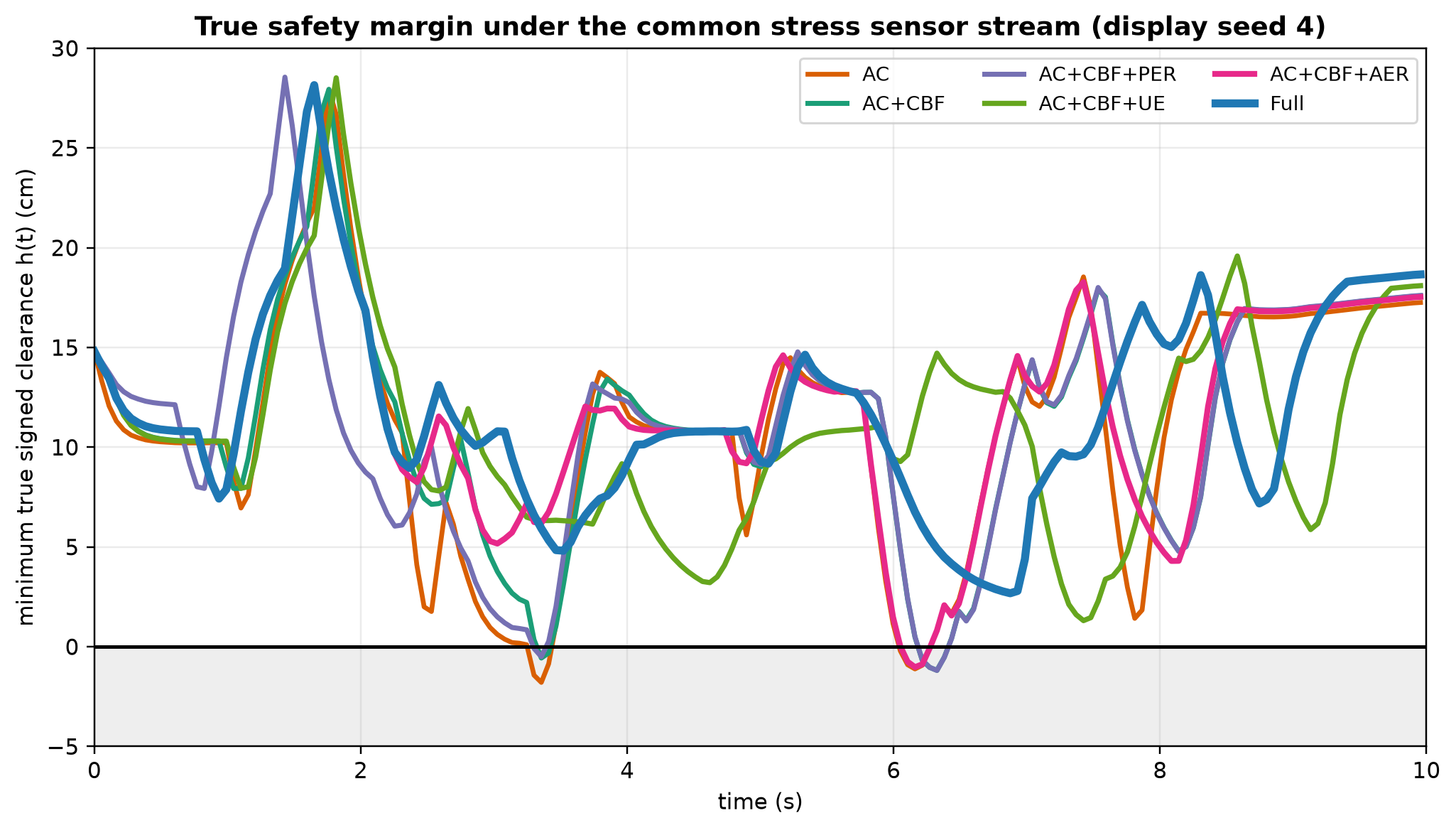}
\caption{True signed clearance over the displayed interval; $h(t)<0$ is unsafe. Reported numerical metrics use the complete registered evaluation horizon.}
\label{fig:safety}
\end{subfigure}
\caption{Extreme-stress trajectories and clearance. Hidden geometry is used only for evaluation.}
\end{figure}

\subsection{Finite-buffer replay audit}

Supplementary Table~S10 separates event generation, finite-buffer retention, and replay sampling. AER and Full retain boundary events strongly, while Full also increases retention and sampling of observable uncertainty events.

\begin{figure}[htbp]
\centering
\includegraphics[width=0.90\textwidth]{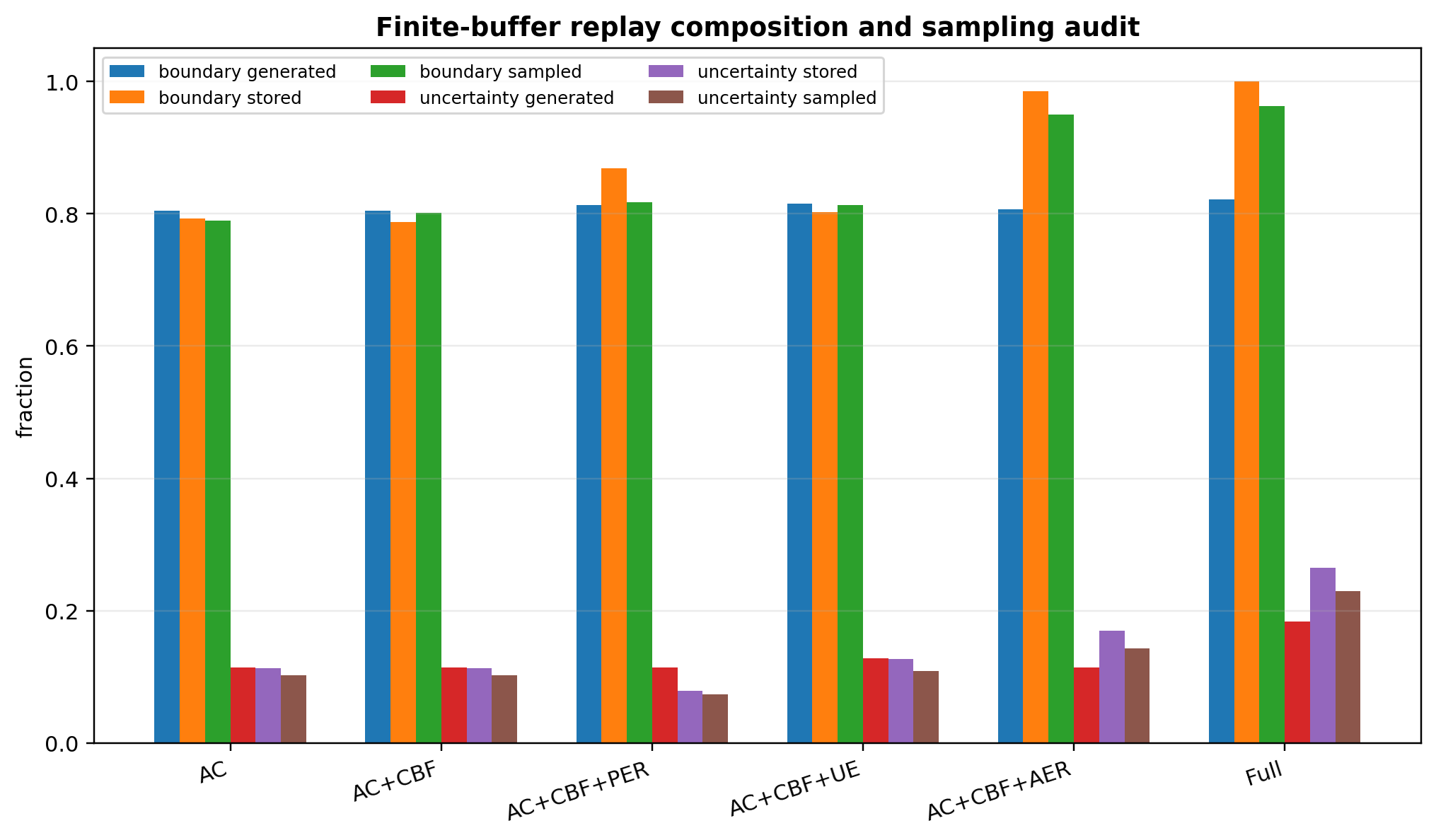}
\caption{Finite-buffer replay composition and sampling; numerical fractions are in Supplementary Table~S10.}
\label{fig:replay}
\end{figure}

\subsection{Moderate test, learning behavior, and robustness sweep}

At multiplier $2.2$, every CBF configuration is collision-free in all five seeds, whereas AC records contact in two. Full has the lowest belief RMSE and the largest clearance; the complete moderate results are in Supplementary Table~S5.

\begin{figure}[htbp]
\centering
\begin{subfigure}[t]{0.49\textwidth}
\includegraphics[width=\textwidth]{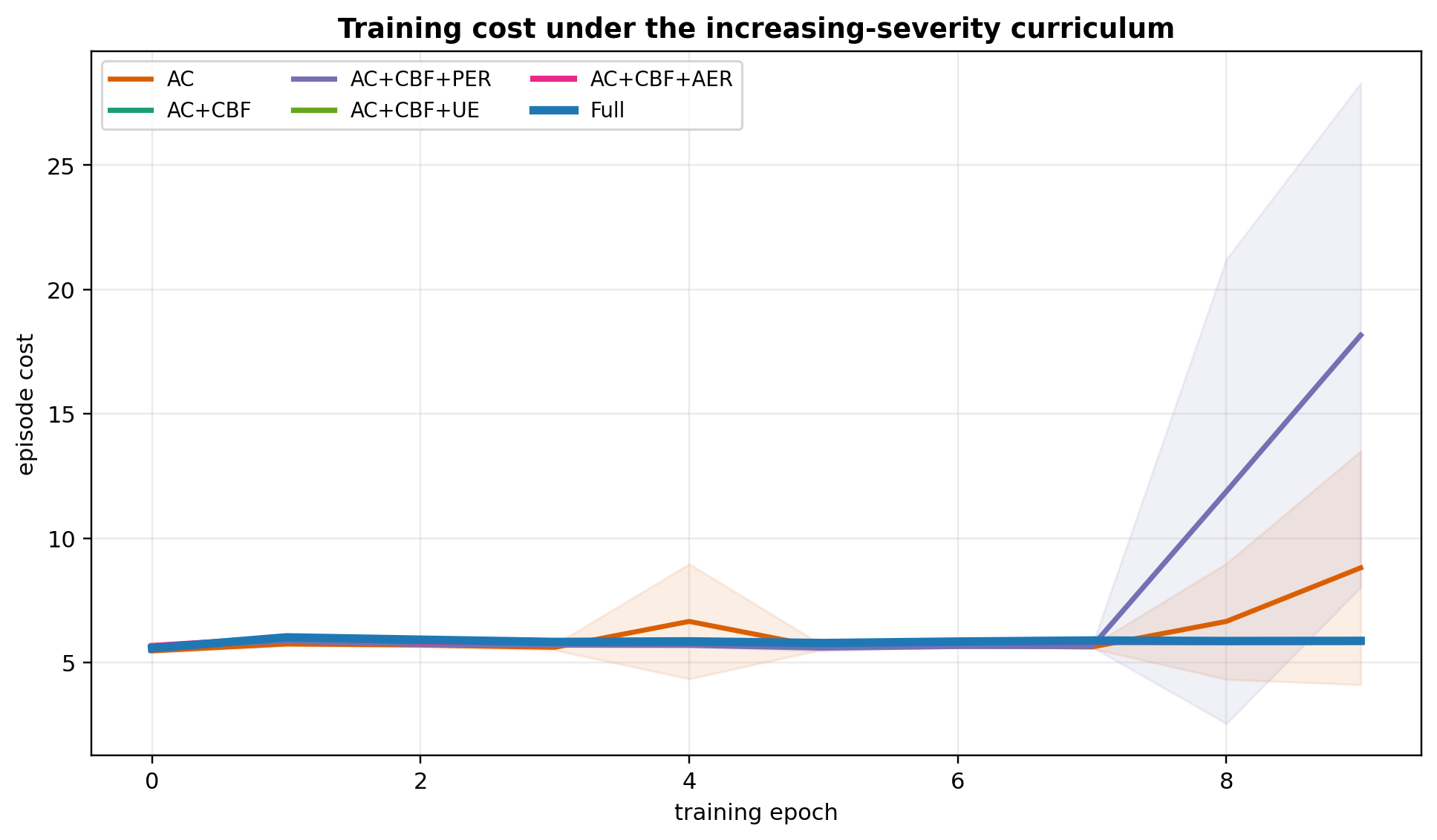}
\caption{Training cost under the increasing-severity curriculum; this is not a fixed-task convergence curve.}
\end{subfigure}\hfill
\begin{subfigure}[t]{0.49\textwidth}
\includegraphics[width=\textwidth]{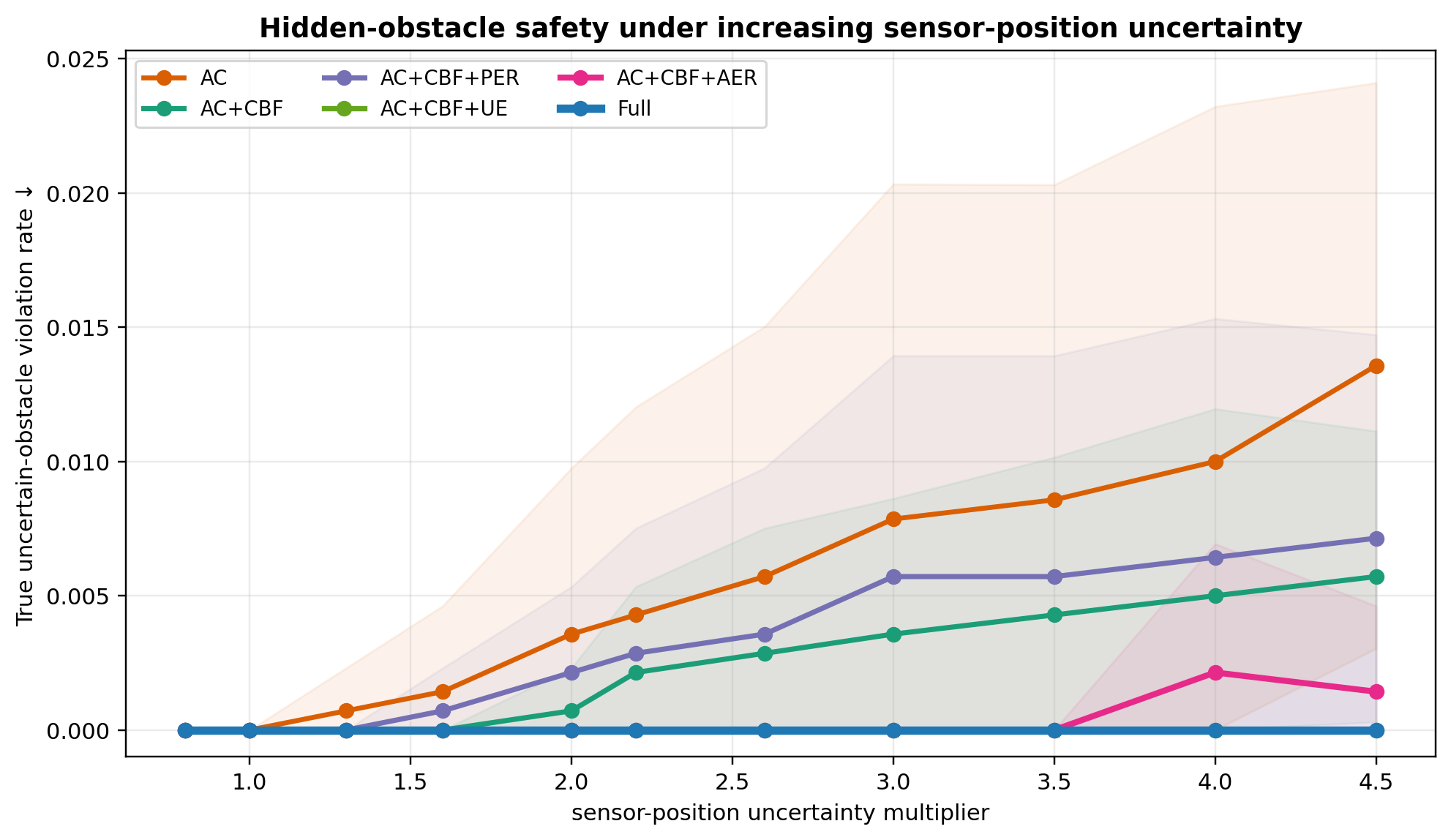}
\caption{Violation rate versus perception severity.}
\end{subfigure}
\caption{Learning and robustness summaries. The sweep uses a fixed episode identifier across severity levels to reduce the severity/random-realization confound.}
\label{fig:learningrobust}
\end{figure}

The sweep and multiplier-$6.0$ test characterize the chosen observation model; they do not define a certified uncertainty radius.

\subsection{Computation and reproducibility}

With $I$ projection sweeps, $J$ active constraints, and action dimension $m=2$, the safety projection costs $O(IJm)$. Uniform replay retrieval is $O(B)$, whereas the present transparent priority implementation scans $N=360$ entries and costs $O(N+B)$ per batch; tree-based sampling could reduce priority operations to $O(\log N)$. Table~\ref{tab:severe} reports machine-specific runtimes.

The executable code, protocol, common streams, seed-level results, and diagnostic tests are supplied as supplementary software at \url{https://github.com/SDNT8810/safe-actor-critic-aer-ue-reproducibility}.
\
\FloatBarrier

\section{Discussion and Application Bridges}
\label{sec:discussion}

Although the integrated viewpoint developed in this paper is intentionally general, it is useful to connect it to recent studies that make its individual components concrete. Such a connection clarifies that the proposed perspective is not speculative; its ingredients already appear in recent works, albeit in partially separated forms. It also helps identify promising future contributions, not by repeating the individual components, but by showing how they can be combined more systematically.

The present perspective is strongly motivated by recent studies that each illuminate one part of the integrated architecture. Safe actor-critic control with experience replay and CBF-based constraints can be formulated meaningfully even when the system is not mechanical. Khalili-Amirabadi et al. \cite{hpv2025} show that safe RL can support public-health intervention design under operational constraints, thereby establishing an application bridge for safety-constrained actor-critic control. Their LSTM-based bi-level control study \cite{lstm2026} demonstrates that online uncertainty estimation and policy optimization can be decomposed into coupled layers for continuous-time nonlinear control. That work supplies the uncertainty-estimation bridge and shows why recurrent architectures may be preferable to memoryless approximators when disturbances are temporally correlated. Moreover, replay should explicitly balance short-term responsiveness and long-horizon diversity through dual-buffer and clustering mechanisms, as shown in the SODACER study \cite{sodacer2026}. That work supplies the adaptive replay bridge. Taken together, these studies motivate the integrated view developed here.

\subsection{What the ablation supports and why integration matters}

The ablation supports three linked observations: corrupted geometry can defeat a nominal CBF, temporal-difference priority does not reliably preserve uncertainty events, and adaptive replay cannot correct a wrong obstacle belief without an estimator. Full combines uncertainty-informed geometry with boundary- and residual-aware replay, yielding zero contacts and five-of-five goal completion in the extreme test. This is a joint benchmark result rather than evidence that every component improves every metric.

\subsection{Robotics, autonomous systems, and health-related design transfer}

The most direct applications are mobile robotics and autonomous systems, where estimator error changes collision geometry and filter interventions identify valuable replay events \cite{hou2025,song2022,ganie2025}. The same interfaces could inform constrained health-control models such as vaccination or treatment allocation \cite{hpv2025}, but this is a design-transfer hypothesis only: the present experiment validates robot navigation, not clinical or public-health performance.

\subsection{Modularity and its cost}

Modularity is an engineering interface, not a plug-and-play convergence theorem. Replacing the static-center filter with an LSTM, Kalman filter, ensemble, or disturbance observer changes the residual distribution, uncertainty calibration, update timing, and valid error bound in the robust CBF inequality. A replacement must expose a documented estimate, calibrated scale, innovation signal, and update rate. Stability and convergence must then be re-established for the selected modules and time scales.

\subsection{Limitations and open challenges}

Despite its appeal, the integrated viewpoint raises several nontrivial challenges. From a theoretical standpoint, combining adaptive replay, recurrent uncertainty estimation, and safety filtering substantially complicates closed-loop analysis. Existing results typically address these components in isolation. A mature theory will likely require separation arguments, incremental-stability tools, or restricted update schedules, in a manner broadly consistent with recent bi-level analyses \cite{lstm2026}.

Computational burden is another important concern. Safety filters, recurrent estimators, and replay clustering all introduce additional overhead. The practical question is whether the resulting gains in safety, robustness, and sample efficiency justify this extra complexity. Cluster maintenance and asynchronous buffer updates, as suggested in recent replay designs, may help mitigate this cost \cite{sodacer2026}.

Replay prioritization also remains application-dependent. Although the mixed-priority form in Eq.~\eqref{eq:genericpriority} is conceptually appealing, selecting appropriate weights and scales is not straightforward. Excessive emphasis on safety-critical transitions may bias the critic toward boundary behavior, whereas insufficient emphasis may weaken safety awareness. Similar considerations apply to uncertainty-related sampling.

Distribution shift and stale data remain persistent concerns in off-policy safe RL. Replayed samples may reflect outdated operating regimes, especially in highly nonstationary environments. In such settings, hybrid offline-online strategies and more robust uncertainty models may be necessary \cite{hickman2025}.

Finally, evaluation protocols should extend beyond reward or cost alone. Meaningful assessment should also include constraint-violation rates, safety margins, estimator residuals, replay diversity, and computational load. Without such a multi-metric evaluation framework, fair comparison among integrated methods remains difficult.

The validation is a two-dimensional simulation with lightweight function approximators, static physical obstacles, five seeds, no independent validation set, and a synthetic extreme detector model selected after initial analysis. The actor update lacks importance correction, the local weight check is post hoc, and the Gaussian disturbance has no hard bound. Proposition~1 concerns replay exposure, not convergence or closed-loop invariance. These limits restrict the evidence to this benchmark and motivate hardware, moving-obstacle, calibrated-sensing, and formal multi-time-scale studies.

\section{Conclusion}
\label{sec:conclusion}

This paper has argued that safe actor-critic optimal control should be viewed as an integrated learning-and-control architecture rather than as a nominal reinforcement-learning algorithm supplemented by a few auxiliary corrections. The central thesis is that adaptive experience replay and online uncertainty estimation are not peripheral implementation details. They shape the data seen by the critic, the reliability of the safety filter, and the responsiveness of the policy under model mismatch. From this standpoint, the proposed perspective contributes three concrete messages. First, replay should preserve safety-critical and uncertainty-informative transitions rather than relying exclusively on uniform or temporal-difference-error-based sampling. Second, uncertainty estimation should be coupled to safe control, because barrier feasibility depends on the true local dynamics. Third, realized transitions produced by the executed safe action, not only the nominal actor output, should drive state-value critic learning and replay formation. The recent literature already contains strong building blocks for this agenda, including CBF-based safe RL, robust adaptive critic control, recurrent uncertainty estimation, clustered or dual-buffer replay, safe HPV control, bi-level LSTM-enhanced nonlinear optimal control, and SODACER-based replay design. These studies make the complementarity of these building blocks particularly visible.

The finite-training bound, robust barrier condition, and component-wise experiment make the proposed interfaces testable. In the extreme test, Full records no contacts, reaches the goal in every seed, and has the lowest mean cost and belief error. This supports the co-design hypothesis on the stated benchmark while leaving convergence, certified safety, and hardware validation open.

%

\section*{Declaration of generative AI and AI-assisted technologies in the manuscript preparation process}

During the preparation of this work, the authors used ChatGPT (OpenAI) solely for language editing to improve English grammar. The authors reviewed and edited the output as needed and take full responsibility for the content of the published article.

\section*{Declaration of competing interest}

The authors declare that they have no known competing financial interests or personal relationships that could have appeared to influence the work reported in this paper.

\clearpage
\setlength{\headheight}{14pt}
\captionsetup{font=small,labelfont=bf}

\setcounter{table}{0}
\renewcommand{\thetable}{S\arabic{table}}
\setcounter{figure}{0}
\renewcommand{\thefigure}{S\arabic{figure}}
\setcounter{equation}{0}
\renewcommand{\theequation}{S\arabic{equation}}

\setlist[itemize]{leftmargin=*,itemsep=.15em,topsep=.25em}
\emergencystretch=2em

\begin{center}
{\LARGE\bfseries Supplementary Material}\par\vspace{2mm}
{\large Toward Integrating Adaptive Experience Replay and Online Uncertainty Estimation in Safe Actor-Critic Optimal Control}\par\vspace{2mm}
Mahshad Rastegarmoghaddam, Davoud Nikkhouy, Shima Samadzadeh
\end{center}

\section{Purpose and reproducibility scope}
This supplement reports the detailed protocol, moderate and exploratory extreme-stress seed-level results, finite-buffer replay audit, fairness tests, information-leakage tests, statistical comparisons, and additional figures supporting the manuscript. All numerical values are generated by the bundled code with 10 training episodes and five seeds. The two uncertain physical obstacles remain static; all six methods receive the same raw noisy detector, exploration, and disturbance streams for a fixed seed and episode. True obstacle centers and true clearance are reserved for simulation evaluation and are excluded from actor, critic, estimator, control-barrier-function (CBF), and replay updates. Multiplier $6.0$ was added after an earlier $4.0$ run and is reported descriptively, not as a preregistered confirmatory test.

\section{Experimental protocol}
\begin{table}[H]
\centering
\caption{Common protocol and implementation settings.}
\label{tab:protocol}
\begin{tabularx}{\textwidth}{>{\bfseries}lX}
\toprule
Setting & Value \\
\midrule
Workspace & $3.20\,\mathrm{m}\times3.20\,\mathrm{m}$ cluttered map with exact rectangular obstacles and two hidden-center circular obstacles.\\
Robot & Radius $0.062\,\mathrm{m}$; planar velocity limit $0.72\,\mathrm{m/s}$; integration step $0.055\,\mathrm{s}$.\\
Start and goal & $(0.22,2.82)\,\mathrm{m}$ and $(1.62,1.58)\,\mathrm{m}$.\\
Uncertain obstacles & Fixed centers $(0.82,1.60)\,\mathrm{m}$ and $(1.82,0.92)\,\mathrm{m}$; radii $0.070\,\mathrm{m}$ and $0.100\,\mathrm{m}$.\\
Training & 10 episodes, 115 steps/episode, five seeds; common perception curriculum $1.0+0.06e$.\\
Evaluation & 280 transitions ending at $15.400\,\mathrm{s}$; last pre-transition sample at $15.345\,\mathrm{s}$; no actor/critic updates.\\
Goal success & Trajectory enters the $5.5$ cm goal tolerance at least once.\\
Tests & Nominal diagnostic $1.0$; moderate post-training evaluation $2.2$; exploratory sweep $0.8$--$4.5$; additional extreme rerun $6.0$; disturbance multiplier $1.0$.\\
Sweep & $0.8,1.0,1.3,1.6,2.0,2.2,2.6,3.0,3.5,4.0,4.5$.\\
Replay & Capacity 360, batch size 44, eight replay updates per training episode.\\
AER/Full weights & $(\lambda_{\mathrm{TD}},\lambda_{\mathrm{safe}},\lambda_{\mathrm{unc}},\lambda_{\mathrm{nov}})=(0.18,0.39,0.30,0.13)$.\\
\bottomrule
\end{tabularx}
\end{table}

The simulator advances the executed closed-loop transition and evaluates the common objective as
\begin{align}
p_{t+1}&=\operatorname{clip}_{\Omega}\!\left[p_t+0.055\,(u_t+d_t)\right],
\qquad \lVert u_t\rVert_2\le0.72,\label{eq:supplant}\\
\ell_t&=0.036\lVert p_t-g\rVert_2^2+0.020\lVert u_t\rVert_2^2
+0.012\lVert u_t-u_{a,t}\rVert_2^2+0.036\lVert u_t-u_{a,t}\rVert_2\notag\\
&\quad{}+5.2\,\mathbb I_{\mathrm{contact},t}
+\max(0,0.010-h_t^{\mathrm{known}})^2 .\label{eq:supcost}
\end{align}
Here $d_t$ is the common exogenous velocity disturbance, $\Omega$ is the robot-center workspace, and the contact bit is an observable bumper event. The hidden true geometry is used to generate physical contact and evaluation metrics, but continuous true clearance, true obstacle centers, true center error, and true gradients are excluded from the replay record and all controller updates.

\begin{table}[H]
\centering
\caption{Exact controller-visible replay schema. The compact generic tuple in the main manuscript is an interface; this table records the implementation-specific fields required to reproduce the replayed critic, actor, estimator, replacement, and sampling updates.}
\label{tab:replayschema}
\small
\begin{tabularx}{\textwidth}{>{\bfseries}p{.20\textwidth}X}
\toprule
Group & Stored fields or meaning \\
\midrule
Executed transition & $p_t,p_{t+1}$; nominal $u_{a,t}$; executed $u_t$; realized instantaneous cost; TD residual and bounded TD score.\\
Replayed actor state & Common Gaussian exploration vector $\epsilon_t=(\epsilon_{x,t},\epsilon_{y,t})$ and active waypoint index $k_t$. These are required by the replayed waypoint update. No behavior-policy likelihood or importance ratio is stored.\\
Safety interface & Estimated barrier slack and active perceived constraint; estimated-barrier gradient; CBF activation; correction $\lVert u_t-u_{a,t}\rVert$; observable contact bit; safety score; boundary-event bit.\\
Uncertainty interface & For each obstacle: raw center $z_{i,t}$, causal belief $\hat c_{i,t}$, posterior scale, detector-reported scale, innovation norm, estimator outlier bit, measurement-validity/hold bit, observable jump score, and known obstacle radius.\\
Replay management & Uncertainty score, state-cell novelty score, uncertainty-event bit, priority, method/configuration identifier, seed, episode, step, time, training flag, severity values, and common sensor-stream hash.\\
Explicitly excluded & True obstacle centers, continuous true clearance, true active constraint and gradient, hidden detector-event label, raw/belief error against truth, and evaluator-only collision-source diagnostics. Evaluation output may contain these fields, but the replay sanitation function cannot copy them.\\
\bottomrule
\end{tabularx}
\end{table}

\subsection{Complete hyperparameters and selection provenance}

No validation-set grid search, random search, Bayesian optimization, or method-specific parameter search was performed. The values below are one common hand-fixed engineering configuration used for the six exploratory ablations; they are not called tuned optima. This removes differential parameter budgets but does not establish that each method is individually optimal. The additional multiplier-$6.0$ condition was introduced after an earlier $4.0$ run. The local replay-weight perturbation in Table~\ref{tab:weightsensitivity} is post hoc and descriptive and was not used to choose or revise the reported weights.

\begingroup
\small
\setlength{\LTleft}{0pt}
\setlength{\LTright}{0pt}
\setlength{\LTcapwidth}{\textwidth}
\begin{longtable}{>{\raggedright\arraybackslash}p{.18\textwidth}>{\raggedright\arraybackslash}p{.28\textwidth}>{\raggedright\arraybackslash}p{.44\textwidth}}
\multicolumn{3}{c}{\parbox{.90\textwidth}{\centering\textbf{Table~\thetable.}\label{tab:hyperparameters} Complete code-grounded implementation constants. Distances are meters and velocities are meters per second unless stated otherwise.}}\\[.6em]
\toprule
Module & Parameter & Fixed value or rule \\
\midrule
\endfirsthead
\multicolumn{3}{l}{\tablename~\thetable\ (continued)}\\
\toprule
Module & Parameter & Fixed value or rule \\
\midrule
\endhead
\midrule
\multicolumn{3}{r}{Continued on next page}\\
\endfoot
\bottomrule
\endlastfoot
Plant & Integration and clipping & Eq.~\eqref{eq:supplant}; $\Delta t=0.055$; $\lVert u\rVert\le0.72$; post-step position clipped to $[r_R,3.20-r_R]^2$, $r_R=0.062$.\\
Plant & Horizons and goal & 115 training transitions; 280 evaluation transitions; goal tolerance $0.055$; success if any of $p_0,\ldots,p_{280}$ enters the tolerance.\\
Objective & Per-transition cost & Eq.~\eqref{eq:supcost}; common to all six rows.\\
Training & Budget & 10 episodes, 8 replay batches per episode, batch size 44, seeds $0,\ldots,4$; training perception $1+0.06e$.\\
Random streams & Integer seed rules & actor $12345+s$; sensor $17000003+100003s+1009e$; disturbance $19001007+200003s+1013e$; exploration $81021+1000s+53e$; replay $202401+101s$.\\
Actor & Initial waypoints & $(0.22,2.82)$, $(0.31,2.43)$, $(0.62,2.25)$, $(0.87,1.99)$, $(0.90,1.62)$, $(1.04,1.27)$, $(1.06,0.82)$, $(1.44,0.70)$, $(2.18,0.70)$, $(2.34,1.14)$, $(1.82,1.39)$, $(1.62,1.58)$.\\
Actor & Initialization/exploration & Internal-waypoint jitter standard deviation $0.006$; Gaussian training exploration standard deviation $0.017$ per velocity component; no exploration in evaluation.\\
Actor & Nominal tracking & waypoint attraction gain $2.25$; terminal-goal gain $0.28$; waypoint switches when current distance $<0.118$ or next distance plus $0.030$ is below current distance.\\
Actor & Replayed policy term & actor learning rate $0.0040$; TD-policy gain $0.18$; cost-TD residual clipped to $[-2.2,2.2]$; normalized exploration clipped componentwise to $[-2.5,2.5]$.\\
Actor & Auxiliary shaping/smoothing & safety, uncertainty, contact pressure weights $(0.58,0.27,0.15)$; waypoint smoothing weights (current, previous, next) $(0.74,0.13,0.13)$.\\
Actor & Waypoint corridor bounds & For internal indices $1{:}10$: $[.24,.38]\!\times\![2.31,2.46]$, $[.53,.77]\!\times\![2.10,2.36]$, $[.77,1.07]\!\times\![1.84,2.14]$, $[.84,1.13]\!\times\![1.52,1.86]$, $[.88,1.11]\!\times\![1.16,1.47]$, $[.97,1.11]\!\times\![.73,.89]$, $[1.29,1.60]\!\times\![.65,.77]$, $[2.02,2.31]\!\times\![.65,.78]$, $[2.16,2.43]\!\times\![1.00,1.26]$, $[1.70,1.94]\!\times\![1.33,1.55]$.\\
Critic & Features & Seven state-value features $[1,g_x,g_y,g_x^2,g_y^2,p_x/3.2,p_y/3.2]$, where $g=p_{\rm goal}-p$.\\
Critic & Bellman update & discount $\gamma=0.985$; learning rate $0.016$; TD clip $[-2.2,2.2]$; normalized increment denominator $1+\lVert\phi\rVert_2^2$.\\
Known potential field & Rectangles/walls & rectangles: robot-radius inflation plus $0.022$, influence $0.18$, gain $0.42$; walls: robot-radius margin plus $0.025$, influence $0.14$, gain $0.34$.\\
Perceived potential field & Circular obstacles & inflation $r_i+r_R+0.010+0.35\sigma_i$; influence $0.31$; gain $0.31$.\\
Estimator & Initialization/reset & learned nominal scale $0.018$; learned gate $0.050$; reset scale floor $0.014$; default causal innovation gate $0.060$.\\
Estimator & Online gain & base gain $0.24$ for UE and $0.18$ for Full; robust weight $\min(1,\mathrm{gate}/\lVert\nu\rVert)$; gain-denominator floor $0.035$; zero gain on invalid measurements.\\
Estimator & Outlier/variance & jump threshold $0.60$ for non-UE observable flags and $0.72$ for UE; variance EMA $0.93/0.07$; posterior $\sigma=\operatorname{clip}(0.008+1.35\sqrt{v}+a,0.014,0.075)$ with outlier addition $a=0.012$ for UE and $0.008$ for Full.\\
Estimator & Full anchor pull & episode reset blend (anchor, first raw) $(0.78,0.22)$; online anchor pull $0.018+0.10\mathbb I_{\rm outlier}$.\\
Estimator & Replay calibration & minimum 8 samples; geometric median (50 iterations maximum, tolerance $10^{-8}$); lower-distance quantile $0.60$; nominal multiplier $1.45$ clipped to $[0.009,0.030]$; gate multiplier $2.7$ clipped to $[0.030,0.058]$; anchor EMA $0.86/0.14$; nominal-scale/gate EMA $0.88/0.12$.\\
CBF & Belief geometry & known inflation $r_R+0.018$; uncertain robust inflation $\rho=0.009+1.20\sigma$; candidate ranges $0.30$ known, $0.40$ uncertain, and $0.26$ walls.\\
CBF & Half-space rule & active when $h<0.20$; class-$\mathcal K$ gain $2.9$; right-side allowance $0.012+0.18\rho$ at physical-disturbance multiplier 1.0. The Gaussian disturbance is not hard bounded.\\
CBF & Projection/guard passes & 4 outer cycles; per cycle 1 all-barrier pass, 2 known-barrier passes, speed clip, then 2 known passes; guard margin $0.0015+0.055(0.012)$ with at most 10 iterations; final 3 known-barrier passes.\\
Replay & Capacity/update & capacity 360; batch 44 without replacement; 8 batches after every training episode; priority floor $10^{-5}$.\\
Replay & Scores/events & score formulas are the main manuscript's TD, safety, uncertainty, and novelty score equations; novelty grid $0.18$; boundary if $\tilde h<0.115$, CBF active, correction $>0.020$, or contact; uncertainty event if score $>0.48$ or estimator outlier.\\
Replay & Priority/sampling & AER/Full weights $(0.18,0.39,0.30,0.13)$ and exponent $\beta=0.78$; PER uses TD-only priority and $\beta=0.68$; uniform rows use equal-probability sampling.\\
Replay & Replacement & uniform rows use reservoir replacement; prioritized rows choose a random replacement candidate with probability $0.16$, otherwise the current minimum-priority entry, and accept if the new priority is no smaller or with fallback probability $0.035$.\\
Replay & Priority refresh & After each critic update, TD scores/priorities are refreshed for the sampled entries only; the whole buffer is not rescored.\\
Sensor & Bias/noise/drift & registration magnitude $U(0.010,0.019)\times$ severity; colored AR coefficient $0.82$ with innovation standard deviation $0.0032\sqrt{\mathrm{severity}}$; white jitter $0.0026\sqrt{\mathrm{severity}}$; sinusoidal amplitudes $(0.009,0.008)\times$ severity.\\
Sensor & Jumps/holds & jump magnitude $U(0.050,0.070)\times$ severity, direction noise $0.22$ rad, decay factor $0.955$, nominal 16--22 affected samples; hold-onset probability $\min(0.012\,\mathrm{severity},0.055)$ and duration 2--4 samples; reported scale $0.010$.\\
Disturbance & Shared process & AR coefficient $0.90$, Gaussian innovation standard deviation $0.0018$ at multiplier 1.0, sinusoidal amplitudes $(0.0040,0.0035)$; stochastic process is not hard bounded.\\
\end{longtable}
\endgroup

\subsection{Post-hoc local replay-weight sensitivity}

A post-hoc local sensitivity diagnostic changes one nominal AER/Full weight by $-20\%$ or $+20\%$, renormalizes the four weights to sum to one, retrains Full with the unchanged five seeds and ten-episode curriculum, and evaluates episode identifier 1001 at multipliers $2.2$ and $6.0$. The nominal rerun reproduces the reported Full seed metrics to within $1.8\times10^{-15}$. This analysis was performed after the reported configuration was fixed and was not used for selection. Across all nine configurations, both tiers remain contact-free and reach the goal in all five seeds. The complete 90 seed rows and unrounded 18 aggregate rows are exported in \texttt{posthoc\_aer\_weight\_sensitivity\_seed\_metrics.csv} and \texttt{posthoc\_aer\_weight\_sensitivity\_aggregate.csv}.

\begin{table}[H]
\centering
\caption{Descriptive one-at-a-time replay-weight sensitivity for Full. Each changed component is multiplied by $0.8$ or $1.2$ and all four weights are then renormalized. Values are mean $\pm$ sample standard deviation over five seeds; $C$ is cost, $H$ is minimum global clearance (cm), and $R$ is belief RMSE (cm). All rows have zero contacts and five successful seeds at both tiers.}
\label{tab:weightsensitivity}
\scriptsize
\begin{adjustbox}{max width=\textwidth}
\begin{tabular}{lccccccl}
\toprule
Configuration & $(\lambda_{\rm TD},\lambda_{\rm safe},\lambda_{\rm unc},\lambda_{\rm nov})$ & $C_{2.2}$ & $H_{2.2}$ & $R_{2.2}$ & $C_{6.0}$ & $H_{6.0}$ & $R_{6.0}$\\
\midrule
Nominal & $(.180,.390,.300,.130)$ & $6.94\pm.30$ & $3.73\pm.36$ & $3.44\pm.22$ & $7.63\pm.44$ & $2.71\pm.23$ & $3.52\pm.55$\\
TD $-20\%$ & $(.149,.405,.311,.135)$ & $6.97\pm.29$ & $3.76\pm.37$ & $3.44\pm.21$ & $7.65\pm.44$ & $2.70\pm.21$ & $3.54\pm.70$\\
TD $+20\%$ & $(.208,.376,.290,.125)$ & $6.99\pm.27$ & $3.77\pm.34$ & $3.44\pm.20$ & $7.66\pm.44$ & $2.68\pm.19$ & $3.50\pm.62$\\
Safety $-20\%$ & $(.195,.338,.325,.141)$ & $6.96\pm.24$ & $3.77\pm.37$ & $3.46\pm.20$ & $7.65\pm.35$ & $2.68\pm.21$ & $3.58\pm.67$\\
Safety $+20\%$ & $(.167,.434,.278,.121)$ & $6.98\pm.29$ & $3.74\pm.40$ & $3.43\pm.22$ & $7.64\pm.42$ & $2.70\pm.22$ & $3.49\pm.65$\\
Uncertainty $-20\%$ & $(.191,.415,.255,.138)$ & $6.99\pm.28$ & $3.78\pm.36$ & $3.45\pm.21$ & $7.66\pm.43$ & $2.71\pm.22$ & $3.55\pm.69$\\
Uncertainty $+20\%$ & $(.170,.368,.340,.123)$ & $6.96\pm.29$ & $3.78\pm.35$ & $3.46\pm.19$ & $7.64\pm.43$ & $2.72\pm.22$ & $3.58\pm.70$\\
Novelty $-20\%$ & $(.185,.400,.308,.107)$ & $6.98\pm.22$ & $3.76\pm.39$ & $3.43\pm.22$ & $7.67\pm.39$ & $2.69\pm.22$ & $3.53\pm.68$\\
Novelty $+20\%$ & $(.175,.380,.292,.152)$ & $6.96\pm.25$ & $3.76\pm.33$ & $3.42\pm.21$ & $7.64\pm.39$ & $2.71\pm.22$ & $3.47\pm.57$\\
\bottomrule
\end{tabular}
\end{adjustbox}
\end{table}

The largest absolute shift from the nominal mean is $0.0490$ in moderate cost, $0.0551$ cm in moderate global clearance, $0.0198$ cm in moderate belief RMSE, $0.0421$ in extreme cost, $0.0277$ cm in extreme global clearance, and $0.0616$ cm in extreme belief RMSE. This local stability is evidence only for the tested $\pm20\%$ neighborhood; it is not a global sensitivity surface, an optimality result, or an independent validation study.

At multiplier $6.0$, the shared detector has persistent registration-bias amplitude $6.0$--$11.4$ cm, smooth drift up to $5.4$ cm in $x$ and $4.8$ cm in $y$, and temporary center jumps of $30$--$42$ cm. Each injected jump has a nominal 16--22-sample ($0.88$--$1.21$ s) decay and scheduled injections can overlap. While no hold is active, the hold-onset probability is $5.5\%$ per step and each onset repeats the prior measurement for two to four frames. The detector nevertheless reports a standard deviation of only $1$ cm, representing a deliberately overconfident sensor under distribution shift. The CBF's $0.012\,\mathrm{m/s}$ disturbance term is a design allowance; the common Gaussian AR(1)-plus-sinusoid disturbance is not hard bounded.

\paragraph{Comparison-method provenance.}
Reference numbers follow the main-manuscript comparison table. AC is the uniform-replay reference configuration [1,2]; AC+CBF is linked to the CBF-safe-control literature [3,4]; AC+CBF+PER adds temporal-difference-prioritized replay [11]; AC+CBF+UE adds online uncertainty estimation [33,20]; AC+CBF+AER adds adaptive safety-aware replay [21]; and Full combines CBF, online uncertainty estimation, and adaptive replay. These labels identify component provenance, not exact reproductions of the cited algorithms. Every numerical value in the following tables was generated by executing the six local configurations on the common benchmark with shared streams and seeds; no numerical result was copied from a cited publication.

\paragraph{Actor-update scope.}
The linear state-value critic is fitted from realized cost and next state produced by the executed post-filter action. Its clipped cost-TD residual directly enters a replay-weighted score-function waypoint update, together with auxiliary controller-visible barrier shaping. Replayed samples contain neither behavior-policy likelihoods nor importance ratios; consequently, this is a biased off-policy surrogate rather than an unbiased on-policy policy-gradient estimator. The benchmark establishes a direct critic-to-actor code path but does not isolate or claim a performance gain caused by the learned value baseline alone.

\section{Moderate post-training results}
\begin{table}[H]
\centering
\caption{Moderate perception-stress results at multiplier $2.2$ (mean $\pm$ standard deviation over five seeds). Main-manuscript reference numbers beside methods identify motivating strands; all values were generated on the common benchmark. For CBF intervention, $\downarrow$ means lower intervention workload, not greater safety.}
\label{tab:moderate}
\scriptsize
\begin{adjustbox}{max width=\textwidth}
\begin{tabular}{lccccccccc}
\toprule
Method & Cost $\downarrow$ & Viol. $\downarrow$ & Min. global $h$ $\uparrow$ & Uncertain clearance $\uparrow$ & Belief RMSE $\downarrow$ & Goal error $\downarrow$ & Goal success $\uparrow$ & CBF int. $\downarrow$ & Runtime $\downarrow$ \\
 & & (\%) & (cm) & (cm) & (cm) & (cm) & (\%) & (\%) & (ms/280-transition episode)\\
\midrule
AC\textsuperscript{[1,2]} & $11.62\pm7.35$ & $0.36\pm0.51$ & $0.03\pm0.36$ & $0.03\pm0.36$ & $6.01\pm0.53$ & $4.67\pm0.81$ & $\mathbf{100.0\pm0.0}$ & N/A & $\mathbf{139.61\pm10.51}$\\
AC+CBF\textsuperscript{[3,4]} & $6.89\pm0.17$ & $\mathbf{0.00\pm0.00}$ & $1.00\pm0.69$ & $1.00\pm0.69$ & $6.01\pm0.53$ & $4.44\pm0.60$ & $\mathbf{100.0\pm0.0}$ & $18.29\pm2.06$ & $229.70\pm12.19$\\
AC+CBF+PER\textsuperscript{[11]} & $\mathbf{6.68\pm0.20}$ & $\mathbf{0.00\pm0.00}$ & $1.41\pm1.00$ & $1.41\pm1.00$ & $6.01\pm0.53$ & $4.43\pm0.61$ & $\mathbf{100.0\pm0.0}$ & $19.07\pm1.93$ & $221.31\pm12.57$\\
AC+CBF+UE\textsuperscript{[33,20]} & $7.36\pm0.32$ & $\mathbf{0.00\pm0.00}$ & $3.23\pm1.13$ & $3.83\pm1.85$ & $5.09\pm0.58$ & $4.52\pm0.66$ & $\mathbf{100.0\pm0.0}$ & $27.29\pm4.57$ & $224.80\pm17.99$\\
AC+CBF+AER\textsuperscript{[21]} & $6.72\pm0.15$ & $\mathbf{0.00\pm0.00}$ & $1.97\pm0.88$ & $1.97\pm0.88$ & $6.01\pm0.53$ & $\mathbf{4.39\pm0.69}$ & $\mathbf{100.0\pm0.0}$ & $\mathbf{18.21\pm1.79}$ & $215.69\pm15.70$\\
Full (present) & $6.94\pm0.30$ & $\mathbf{0.00\pm0.00}$ & $\mathbf{3.73\pm0.36}$ & $\mathbf{4.55\pm1.26}$ & $\mathbf{3.44\pm0.22}$ & $4.40\pm0.75$ & $\mathbf{100.0\pm0.0}$ & $21.50\pm3.27$ & $229.91\pm10.60$\\
\bottomrule
\end{tabular}
\end{adjustbox}
\end{table}

\section{Additional exploratory extreme-stress results}
\begin{table}[H]
\centering
\caption{Additional exploratory extreme-stress results at multiplier $6.0$ (mean $\pm$ sample standard deviation over five seeds). Main-manuscript reference numbers beside methods identify motivating strands; all values were generated on the common benchmark. For CBF intervention, $\downarrow$ means lower intervention workload, not greater safety.}
\label{tab:extreme}
\scriptsize
\begin{adjustbox}{max width=\textwidth}
\begin{tabular}{lccccccccc}
\toprule
Method & Cost $\downarrow$ & Viol. $\downarrow$ & Min. global $h$ $\uparrow$ & Uncertain clearance $\uparrow$ & Belief RMSE $\downarrow$ & Goal error $\downarrow$ & Goal success $\uparrow$ & CBF int. $\downarrow$ & Runtime $\downarrow$ \\
 & & (\%) & (cm) & (cm) & (cm) & (cm) & (\%) & (\%) & (ms/280-transition episode)\\
\midrule
AC\textsuperscript{[1,2]} & $46.99\pm19.28$ & $2.79\pm1.32$ & $-1.76\pm0.80$ & $-1.76\pm0.80$ & $16.71\pm1.55$ & $\mathbf{4.00\pm0.69}$ & $\mathbf{100.0\pm0.0}$ & N/A & $\mathbf{142.03\pm13.87}$\\
AC+CBF\textsuperscript{[3,4]} & $31.00\pm12.43$ & $1.64\pm0.86$ & $-1.26\pm0.40$ & $-1.26\pm0.40$ & $16.71\pm1.55$ & $4.62\pm0.69$ & $\mathbf{100.0\pm0.0}$ & $19.79\pm4.38$ & $223.44\pm10.32$\\
AC+CBF+PER\textsuperscript{[11]} & $39.08\pm24.08$ & $2.21\pm1.64$ & $-1.04\pm1.91$ & $-1.04\pm1.91$ & $16.71\pm1.55$ & $4.47\pm0.76$ & $\mathbf{100.0\pm0.0}$ & $\mathbf{18.36\pm2.38}$ & $220.59\pm15.82$\\
AC+CBF+UE\textsuperscript{[33,20]} & $8.96\pm2.08$ & $\mathbf{0.00\pm0.00}$ & $\mathbf{3.22\pm2.81}$ & $3.94\pm3.37$ & $11.08\pm1.23$ & $17.72\pm29.14$ & $80.0\pm44.7$ & $47.64\pm28.19$ & $236.34\pm17.71$\\
AC+CBF+AER\textsuperscript{[21]} & $22.62\pm12.82$ & $1.07\pm0.87$ & $-0.35\pm1.55$ & $-0.35\pm1.55$ & $16.71\pm1.55$ & $4.37\pm0.67$ & $\mathbf{100.0\pm0.0}$ & $21.64\pm3.66$ & $217.80\pm12.77$\\
Full (present) & $\mathbf{7.63\pm0.44}$ & $\mathbf{0.00\pm0.00}$ & $2.71\pm0.23$ & $\mathbf{6.95\pm0.26}$ & $\mathbf{3.52\pm0.55}$ & $4.70\pm0.51$ & $\mathbf{100.0\pm0.0}$ & $29.14\pm4.69$ & $228.76\pm15.32$\\
\bottomrule
\end{tabular}
\end{adjustbox}
\end{table}

\subsection{Seed-level exploratory extreme-stress outcomes}
\scriptsize
\begin{longtable}{llrrrrrr}
\caption{Seed-level outcomes in the additional multiplier-$6.0$ rerun.}\label{tab:seedlevel}\\
\toprule Method & Seed & Cost & Viol. count & Min. $h$ (cm) & Belief RMSE (cm) & Goal error (cm) & Reached\\
\midrule
\endfirsthead
\toprule Method & Seed & Cost & Viol. count & Min. $h$ (cm) & Belief RMSE (cm) & Goal error (cm) & Reached\\
\midrule
\endhead
AC & 0 & 42.80 & 7 & -1.12 & 17.03 & 3.30 & yes \\
AC & 1 & 63.49 & 11 & -2.65 & 17.14 & 4.61 & yes \\
AC & 2 & 16.80 & 2 & -0.80 & 17.87 & 3.42 & yes \\
AC & 3 & 63.86 & 11 & -2.43 & 14.00 & 3.86 & yes \\
AC & 4 & 48.01 & 8 & -1.79 & 17.50 & 4.81 & yes \\
AC\allowbreak+CBF & 0 & 27.98 & 4 & -1.14 & 17.03 & 5.20 & yes \\
AC\allowbreak+CBF & 1 & 48.60 & 8 & -1.81 & 17.14 & 4.21 & yes \\
AC\allowbreak+CBF & 2 & 17.24 & 2 & -0.73 & 17.87 & 3.84 & yes \\
AC\allowbreak+CBF & 3 & 23.19 & 3 & -1.44 & 14.00 & 4.39 & yes \\
AC\allowbreak+CBF & 4 & 37.99 & 6 & -1.18 & 17.50 & 5.49 & yes \\
AC\allowbreak+CBF\allowbreak+PER & 0 & 38.21 & 6 & -1.29 & 17.03 & 4.87 & yes \\
AC\allowbreak+CBF\allowbreak+PER & 1 & 74.49 & 13 & -2.81 & 17.14 & 4.21 & yes \\
AC\allowbreak+CBF\allowbreak+PER & 2 & 6.49 & 0 & 2.16 & 17.87 & 3.42 & yes \\
AC\allowbreak+CBF\allowbreak+PER & 3 & 38.59 & 6 & -2.09 & 14.00 & 4.37 & yes \\
AC\allowbreak+CBF\allowbreak+PER & 4 & 37.64 & 6 & -1.19 & 17.50 & 5.47 & yes \\
AC\allowbreak+CBF\allowbreak+UE & 0 & 7.44 & 0 & 0.66 & 12.17 & 5.06 & yes \\
AC\allowbreak+CBF\allowbreak+UE & 1 & 11.13 & 0 & 7.83 & 12.60 & 69.83 & no \\
AC\allowbreak+CBF\allowbreak+UE & 2 & 7.30 & 0 & 3.05 & 10.30 & 3.61 & yes \\
AC\allowbreak+CBF\allowbreak+UE & 3 & 11.34 & 0 & 3.23 & 9.82 & 5.15 & yes \\
AC\allowbreak+CBF\allowbreak+UE & 4 & 7.60 & 0 & 1.31 & 10.51 & 4.96 & yes \\
AC\allowbreak+CBF\allowbreak+AER & 0 & 27.82 & 4 & -1.16 & 17.03 & 5.09 & yes \\
AC\allowbreak+CBF\allowbreak+AER & 1 & 38.54 & 6 & -1.65 & 17.14 & 4.25 & yes \\
AC\allowbreak+CBF\allowbreak+AER & 2 & 6.71 & 0 & 2.26 & 17.87 & 3.37 & yes \\
AC\allowbreak+CBF\allowbreak+AER & 3 & 12.59 & 1 & -0.17 & 14.00 & 4.26 & yes \\
AC\allowbreak+CBF\allowbreak+AER & 4 & 27.43 & 4 & -1.02 & 17.50 & 4.88 & yes \\
Full & 0 & 7.55 & 0 & 2.64 & 3.07 & 5.32 & yes \\
Full & 1 & 7.48 & 0 & 2.90 & 3.92 & 4.44 & yes \\
Full & 2 & 7.21 & 0 & 2.94 & 3.89 & 3.99 & yes \\
Full & 3 & 8.38 & 0 & 2.37 & 2.77 & 4.74 & yes \\
Full & 4 & 7.53 & 0 & 2.68 & 3.94 & 4.98 & yes \\
\bottomrule
\end{longtable}

\section{Statistical comparisons}
The omnibus Friedman tests across the six paired methods give $\chi_F^2=17.00$ ($p=0.00450$) for cost, $\chi_F^2=21.23$ ($p=7.32\times10^{-4}$) for violation rate, $\chi_F^2=20.43$ ($p=0.00104$) for minimum clearance, and $\chi_F^2=25.00$ ($p=1.39\times10^{-4}$) for belief RMSE. Goal error is not significant ($\chi_F^2=10.37$, $p=0.0654$). For CBF intervention among the five CBF methods, $\chi_F^2=18.40$ and $p=0.00103$. Because only five pairs are available, the two-sided Wilcoxon tests have coarse resolution and are reported as exploratory paired comparisons alongside the seed-level outcomes; they do not establish statistical superiority. All $p$-values are unadjusted across the reported metrics and pairs, and no multiplicity-adjusted confirmatory inference is made. Both test families and the SciPy version are exported by the bundled software.

\begin{table}[H]
\centering
\caption{Complete exported Friedman tests for the additional extreme-stress evaluation. The CBF-intervention row excludes AC because intervention is not defined without a CBF. Values are unadjusted exploratory diagnostics.}
\label{tab:friedman}
\small
\begin{tabular}{llrrr}
\toprule
Scope & Metric & Methods & $\chi_F^2$ & $p$ value\\
\midrule
All six & Cost & 6 & 17.0000 & 0.0044998\\
All six & Violation rate & 6 & 21.2342 & 0.0007315\\
All six & Minimum global clearance & 6 & 20.4286 & 0.0010382\\
All six & Belief-center RMSE & 6 & 25.0000 & 0.0001393\\
All six & Goal error & 6 & 10.3714 & 0.0653696\\
Five CBF rows & CBF intervention rate & 5 & 18.4000 & 0.0010306\\
\bottomrule
\end{tabular}
\end{table}

\begin{table}[H]
\centering
\caption{Two-sided paired Wilcoxon comparisons against Full in the additional extreme-stress test. With five pairs and ties, the results are exploratory; $p$-values are unadjusted.}
\label{tab:wilcoxon}
\small
\begin{tabular}{llcc}
\toprule
Comparison & Metric & Statistic & $p$ value\\
\midrule
AC vs Full & cost / violation / min clearance / belief RMSE & $0/0/0/0$ & $0.0625/0.0625/0.0625/0.0625$\\
AC+CBF vs Full & cost / violation / min clearance / belief RMSE & $0/0/0/0$ & $0.0625/0.0625/0.0625/0.0625$\\
AC+CBF+PER vs Full & cost / violation / min clearance / belief RMSE & $1/0/0/0$ & $0.125/0.125/0.0625/0.0625$\\
AC+CBF+UE vs Full & cost / violation / min clearance / belief RMSE & $3/0/7/0$ & $0.3125/1.000/1.000/0.0625$\\
AC+CBF+AER vs Full & cost / violation / min clearance / belief RMSE & $1/0/0/0$ & $0.125/0.125/0.0625/0.0625$\\
\bottomrule
\end{tabular}
\end{table}

\section{Finite-buffer replay audit}
\begin{table}[H]
\centering
\caption{Generated, retained, and sampled event fractions (mean $\pm$ standard deviation over five seeds). Main-manuscript method-reference mapping is AC [1,2], AC+CBF [3,4], AC+CBF+PER [11], AC+CBF+UE [33,20], AC+CBF+AER [21], and Full (present).}
\label{tab:replayaudit}
\scriptsize
\begin{adjustbox}{max width=\textwidth}
\begin{tabular}{lcccccc}
\toprule
Method & Generated boundary & Stored boundary & Sampled boundary & Generated uncertainty & Stored uncertainty & Sampled uncertainty\\
\midrule
AC & $0.804\pm0.004$ & $0.793\pm0.025$ & $0.790\pm0.019$ & $0.115\pm0.016$ & $0.113\pm0.010$ & $0.102\pm0.015$\\
AC+CBF & $0.805\pm0.004$ & $0.788\pm0.027$ & $0.802\pm0.016$ & $0.115\pm0.016$ & $0.113\pm0.010$ & $0.102\pm0.015$\\
AC+CBF+PER & $0.813\pm0.004$ & $0.868\pm0.009$ & $0.818\pm0.009$ & $0.115\pm0.016$ & $0.079\pm0.021$ & $0.073\pm0.015$\\
AC+CBF+UE & $0.815\pm0.002$ & $0.803\pm0.029$ & $0.813\pm0.013$ & $0.128\pm0.016$ & $0.127\pm0.018$ & $0.108\pm0.020$\\
AC+CBF+AER & $0.807\pm0.005$ & $0.985\pm0.002$ & $0.950\pm0.005$ & $0.115\pm0.016$ & $0.170\pm0.030$ & $0.143\pm0.026$\\
Full & $0.821\pm0.004$ & $1.000\pm0.000$ & $0.963\pm0.004$ & $0.184\pm0.024$ & $0.264\pm0.026$ & $0.229\pm0.036$\\
\bottomrule
\end{tabular}
\end{adjustbox}
\end{table}


\section{Fairness, no-oracle, and collision-source audits}
\begin{itemize}
\item \textbf{Shared sensor stream:} for all 15 nominal/moderate/extreme seed-evaluation groups, six methods are present, there is exactly one within-run composite sensor-stream hash covering raw centers, validity bits, jump scores, and reported scales, and the across-method ranges of raw sensor RMSE, hold fraction, and jump mean are all zero. The hash certifies within-run equality; it is not claimed to be portable across arbitrary numerical-library platforms.
\item \textbf{Hidden-truth invariance:} translating hidden true centers by $(0.43,-0.31)$ m while keeping all controller-visible signals fixed gives maximum nominal- and executed-action differences of zero for every method.
\item \textbf{Known-map collisions:} the audit reports known-wall/rectangle violations separately from uncertain-obstacle contacts in the additional extreme test.
\item \textbf{Replay sanitation:} true obstacle center, true signed clearance, true center error, and true gradients are absent from stored learning transitions.
\end{itemize}

\section{Additional figures}


\begin{figure}[H]
\centering
\begin{subfigure}[t]{.49\textwidth}\includegraphics[width=\textwidth]{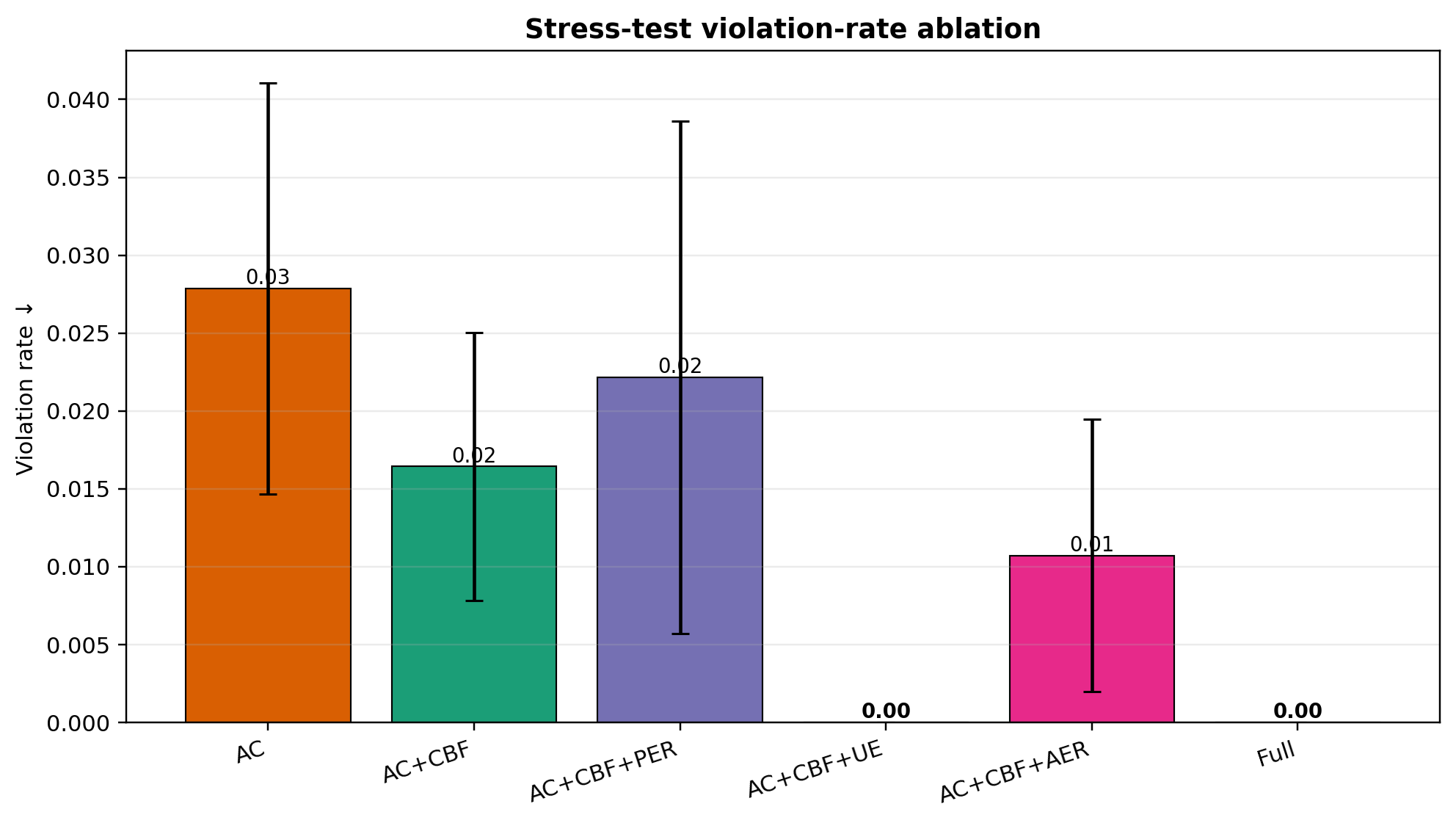}\caption{Violation ablation.}\end{subfigure}\hfill
\begin{subfigure}[t]{.49\textwidth}\includegraphics[width=\textwidth]{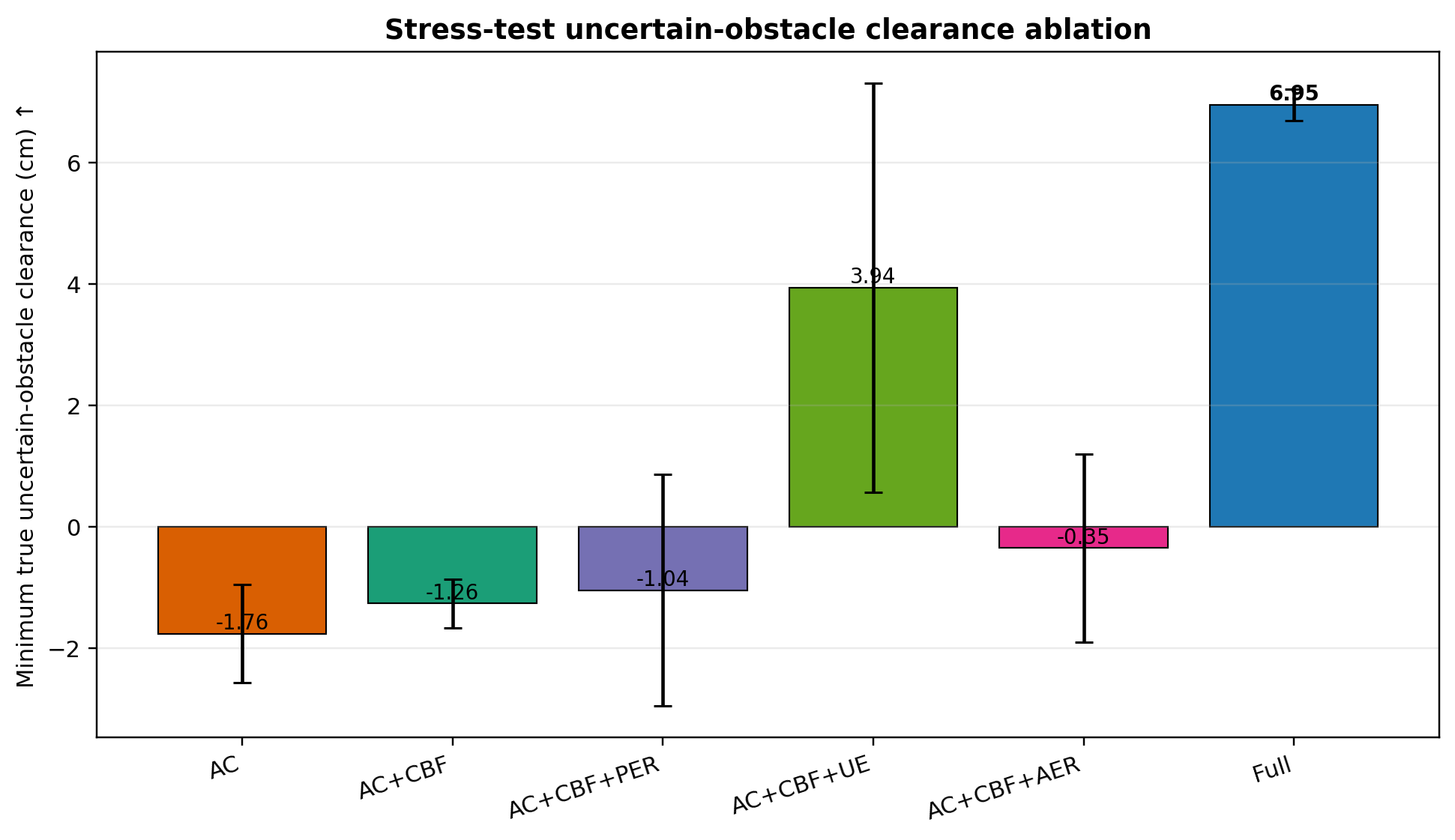}\caption{Minimum uncertain-obstacle clearance.}\end{subfigure}
\caption{Additional extreme-stress safety summaries. Bars show means and error bars show sample standard deviations over five seeds; individual seed points are not overlaid.}
\end{figure}

\begin{figure}[H]
\centering
\begin{subfigure}[t]{.49\textwidth}\includegraphics[width=\textwidth]{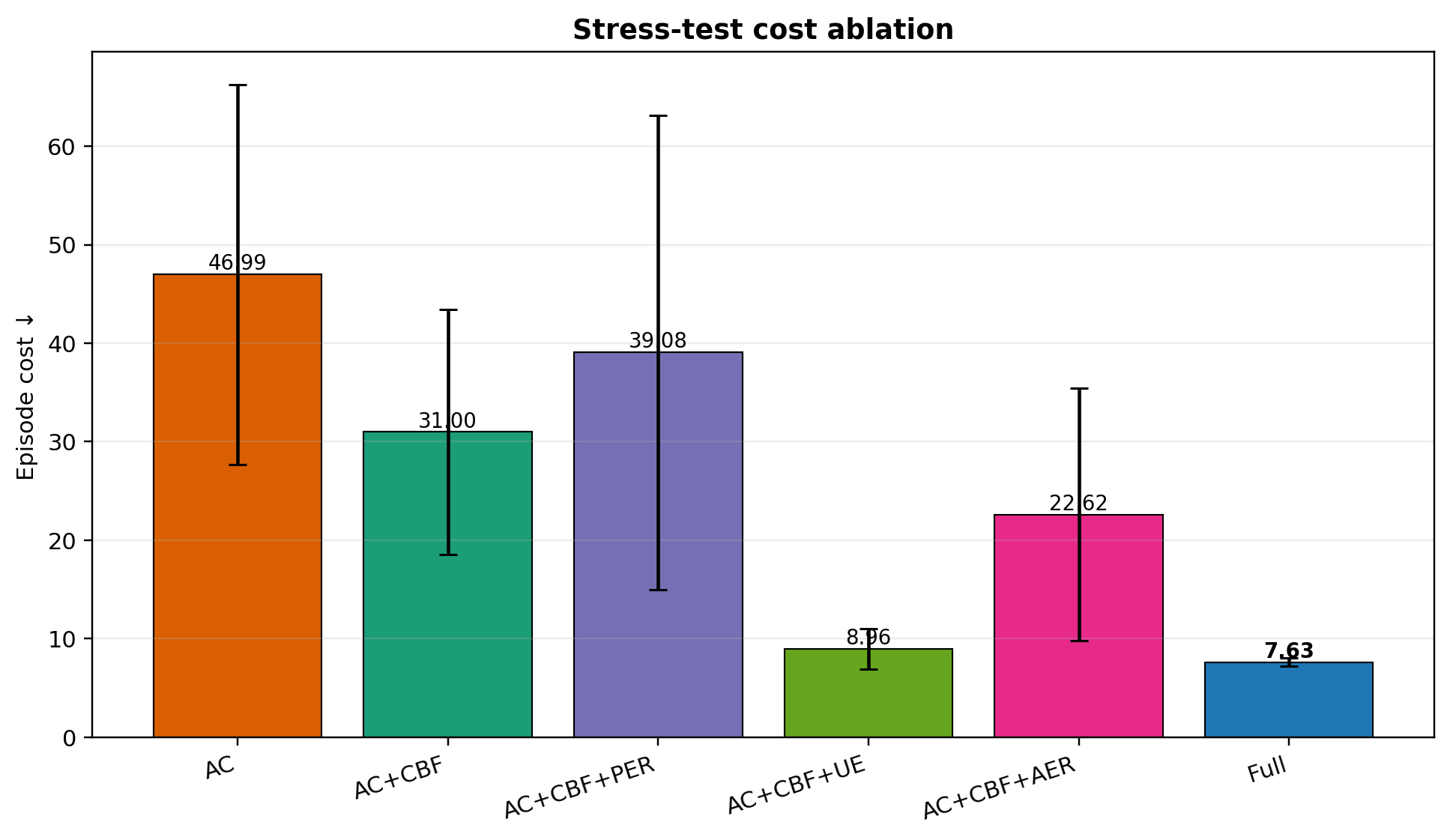}\caption{Cost.}\end{subfigure}\hfill
\begin{subfigure}[t]{.49\textwidth}\includegraphics[width=\textwidth]{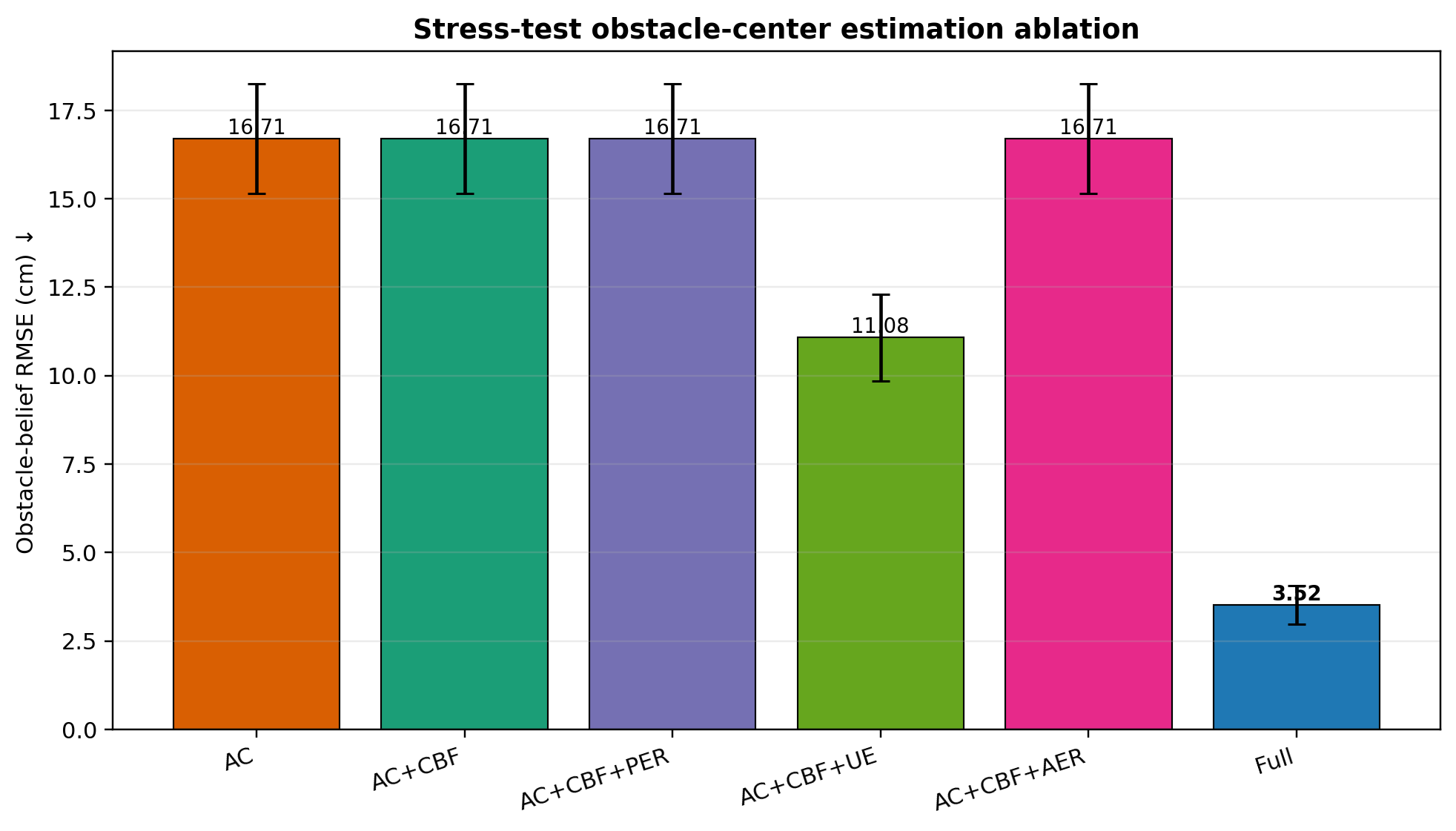}\caption{Belief RMSE.}\end{subfigure}
\caption{Exploratory extreme-stress performance and estimation summaries; bars show mean $\pm$ sample standard deviation over five seeds.}
\end{figure}

\begin{figure}[H]
\centering
\begin{subfigure}[t]{.49\textwidth}\includegraphics[width=\textwidth]{fig09_robustness_sweep_violation.png}\caption{Violation rate.}\end{subfigure}\hfill
\begin{subfigure}[t]{.49\textwidth}\includegraphics[width=\textwidth]{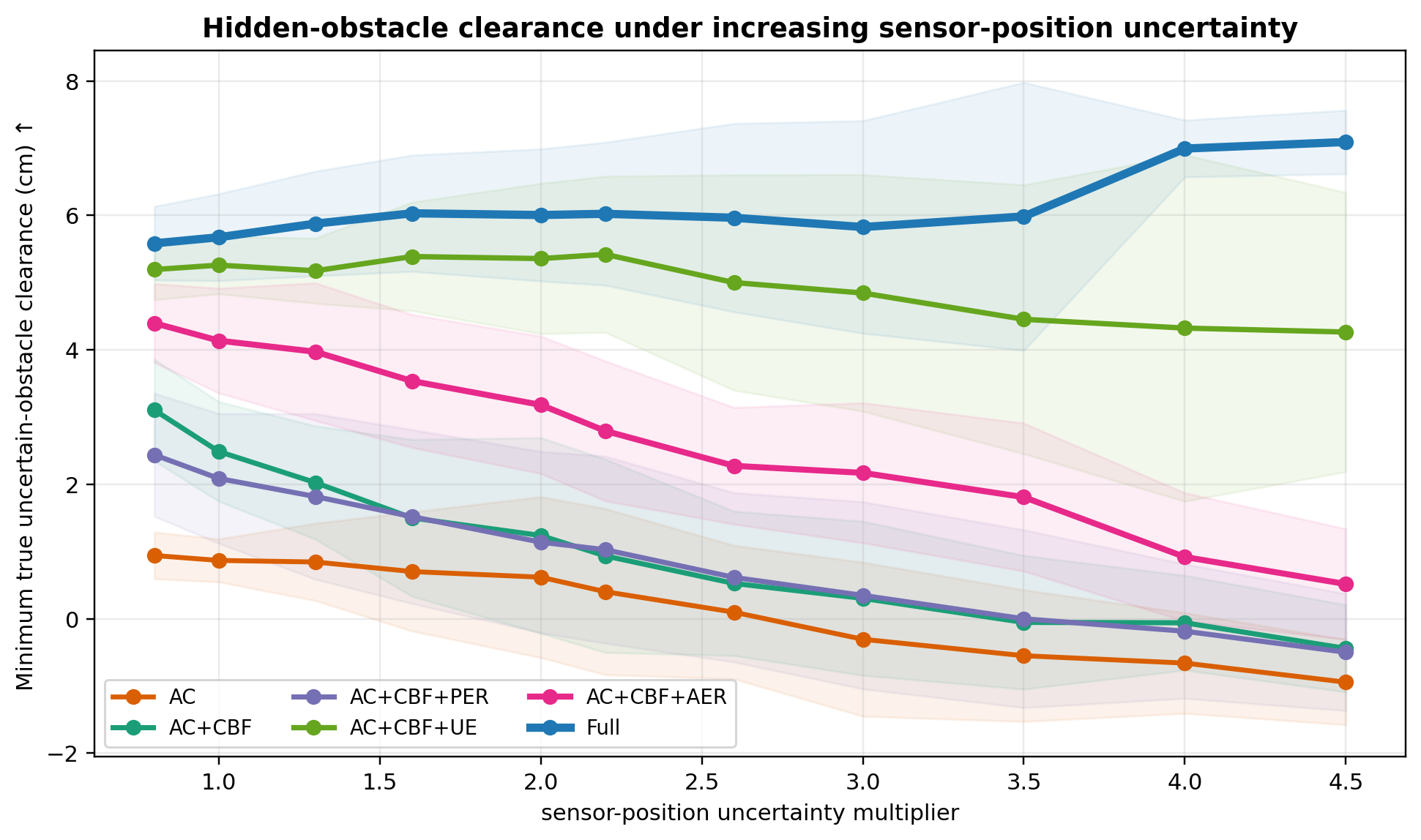}\caption{Minimum true clearance.}\end{subfigure}
\caption{Eleven-level perception-uncertainty sweep.}
\end{figure}

\begin{figure}[H]
\centering
\includegraphics[width=.65\textwidth]{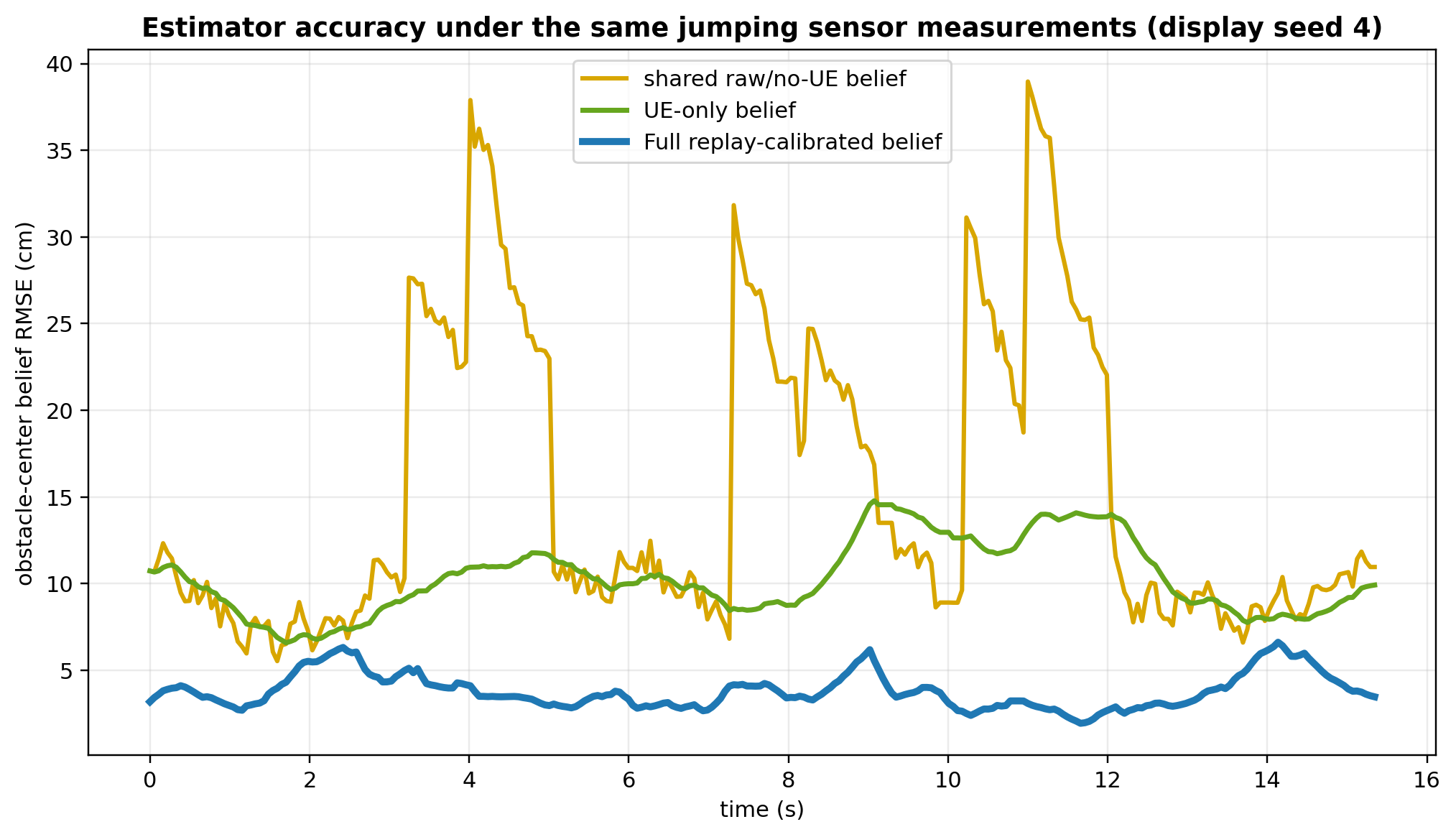}
\caption{Obstacle-center belief error over time in the severe visualization episode.}
\end{figure}

\begin{figure}[H]
\centering
\includegraphics[width=.97\textwidth]{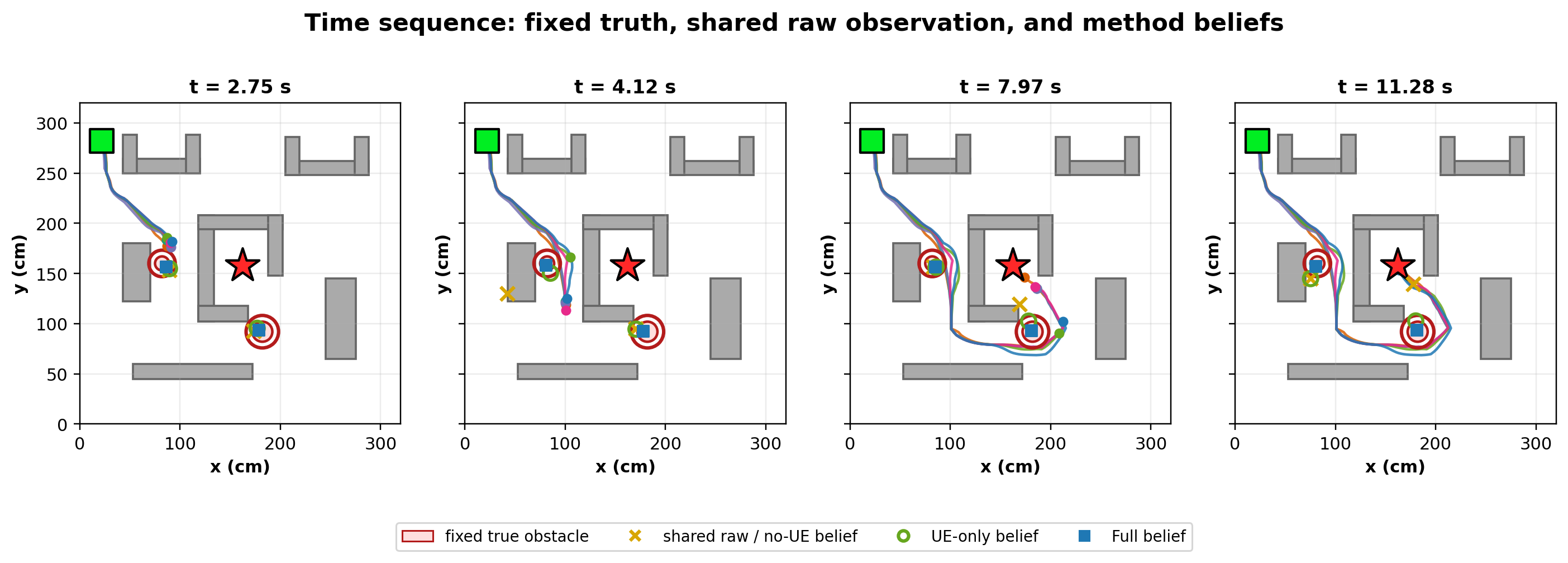}
\caption{Time-sequence panels illustrating fixed physical obstacles and changing detector/belief centers.}
\end{figure}


\end{document}